\documentclass[tighten,iop,apj]{emulateapj}

\usepackage{amsmath}
\usepackage{graphicx}
\usepackage{url}
\usepackage{color}

\newcommand{\ud}{\ensuremath{\mathrm{d}}}

\begin{document}

\title{SASI Activity in Three-Dimensional Neutrino-Hydrodynamics 
Simulations of Supernova Cores}

\shorttitle{SASI in 3D Supernova Models with Neutrino Transport}
\shortauthors{Hanke et al.}
\author{Florian~Hanke, Bernhard~M\"uller, Annop~Wongwathanarat,
        Andreas~Marek, Hans-Thomas~Janka}
\affil{Max-Planck-Institut f\"ur Astrophysik, \
       Karl-Schwarzschild-Str. 1, 85748 Garching, Germany; \
       \{fhanke,bjmuellr,annop,amarek,thj\}@mpa-garching.mpg.de}

\begin{abstract}
The relevance of the standing accretion shock instability (SASI)
compared to neutrino-driven convection in three-dimensional (3D)
supernova-core environments is still highly controversial.
Studying a 27\,$M_\odot$ progenitor,
we demonstrate, for the first time, that violent
SASI activity can develop in 3D simulations with detailed neutrino 
transport despite the presence of
convection. This result was obtained with the
\textsc{Prometheus-Vertex} code with the same
sophisticated neutrino treatment so far used only in
1D and 2D models. While buoyant plumes initially determine the nonradial 
mass motions in the postshock layer, bipolar shock sloshing with
growing amplitude sets in during a phase
of shock retraction and turns into a violent spiral mode whose
growth is only quenched when the infall of the Si/SiO 
interface leads to strong shock expansion in response to a 
dramatic decrease of the mass accretion rate. In the phase
of large-amplitude SASI sloshing and spiral motions, the postshock 
layer exhibits nonradial deformation dominated by the
lowest-order spherical harmonics ($\ell = 1$, $m=0,\pm 1$)
in distinct contrast to the higher multipole structures associated with
neutrino-driven convection. We find that the SASI amplitudes, shock
asymmetry, and nonradial kinetic energy in 3D can exceed those of
the corresponding 2D case during extended periods of the
evolution. We also perform parametrized 3D simulations of a
25\,$M_\odot$ progenitor, using a simplified, gray neutrino transport
scheme, an axis-free Yin-Yang grid, and different amplitudes of 
random seed perturbations. They confirm the importance of
the SASI for another progenitor, its independence of the
choice of spherical grid, and its preferred growth for
fast accretion flows connected to small shock radii and compact 
proto-neutron stars as previously found in 2D setups.
\end{abstract}


\section{Introduction}
\label{sec:intro}
Today there is a universal consensus in supernova theory that the
neutrino-driven explosion mechanism hinges vitally on the supporting
action of multidimensional hydrodynamic instabilities (with the
notable exception of low-mass progenitors with O-Ne-Mg core;
\citealp{kitaura_06}). Two different hydrodynamical instabilities can
potentially operate in the accretion flow between the stalled
supernova shock and the forming neutron star: On the one hand,
neutrino heating establishes a negative entropy gradient in the gain
layer, which is unstable to buoyancy-driven convection, a fact that
has long been recognized \citep{bethe_90} and was confirmed already by
the first generation of multi-dimensional simulations
\citep{herant_92,burrows_92,herant_94,burrows_95,janka_96,mueller_97}.
Later, purely hydrodynamical simulations of adiabatic accretion flows
in the downstream region of a stalled shock \citep{blondin_03}
revealed the existence of a distinctly different phenomenon, the
``standing-accretion shock instability'' (SASI), which involves
large-scale $\ell=1$ and $\ell=2$ sloshing motions of the shock front
and in three dimensions can also develop $\ell=1, m = \pm 1$ spiral
modes \citep{blondin_07,iwakami_09,fernandez_10}.  The characteristic
oscillatory growth of this instability is thought to be mediated by an
amplifying advective-acoustic cycle
(\citealp{foglizzo_02,foglizzo_06,foglizzo_07,guilet_12};
see, however, \citealt{blondin_06} and \citealt{laming_07} for the
interpretation as a purely acoustic cycle).  The hypothesis of an
advective-acoustic process receives strong support by a detailed
analysis of the mode frequencies \citep{scheck_08,guilet_12} and by a
shallow water analogue, the SWASI experiment, in which the water flow
between a circular reservoir and a central tube as sink mimics the
accretion flow that feeds the nascent neutron star
\citep{foglizzo_12}. With the role of the shock and of acoustic waves
being played by a hydraulic jump and surface water waves,
respectively, similar sloshing and spiral motions are observed as in
the hydrodynamical simulations of collapsing stellar cores.

Violent hydrodynamic mass motions in the postshock layer do not only
have the potential to improve the heating conditions in the supernova
core by pushing the shock farther out and by prolonging the dwell time
of the accreted matter in the heating region, which has supportive
consequences for the supernova explosion
\citep{herant_94,burrows_95,janka_96,fryer_00,marek_09,murphy_08,
  nordhaus_10,hanke_12,suwa_10}.  Convection and the SASI also create
seed asymmetries that determine the ejecta morphology during the
explosion phase \citep{kifonidis_03,kifonidis_06}. Moreover,
large-scale anisotropies in the ejecta may yield a natural explanation
for the observed high kick velocities of young pulsars
\citep{janka_94,herant_95,burrows_96,scheck_04,scheck_06,wongwathanarat_10,nordhaus_10b,nordhaus_12,wongwathanarat_12}. The
$\ell=1,|m|=1$ spiral mode of the SASI could also provide an effective
means for spinning up the proto-neutron star (PNS) \citep{blondin_07}.
Moreover, large-amplitude SASI motions were found to stir strong
g-mode activity in the neutron star surface, which leads to
gravitational-wave emission \citep{marek_08,murphy_09}.
Quasi-periodic variations of the accretion flow onto the neutron star
due to SASI shock oscillations may also cause fluctuations of the
radiated neutrino luminosities and mean energies
\citep{marek_08,ott_08_a,brandt_11}. The corresponding neutrino-signal
fluctuations on millisecond timescales could be measured for a
galactic supernova with good time resolution as, e.g., provided by the
IceCube detector \citep{lund_10,lund_12} and could serve as valuable
probe of neutrino properties \citep{ellis_12a,ellis_12b} as well as of
the dynamics in the supernova core.

Up to now most investigations of convection and the SASI in
core-collapse supernovae have relied on axisymmetric (2D) simulations.
Three-dimensional (3D) models based on various approximations for
treating neutrino heating and cooling in the supernova core have only
recently become available, but have already sparked a controversy
about the development and mutual interaction of the two instabilities
in 3D. \citet{burrows_12} and \citet{dolence_13}, who conducted
simulations using a simple light-bulb neutrino
scheme, were rather outspoken in
classifying the SASI as a subdominant phenomenon in the presence of
neutrino heating. They argued that the violent sloshing motions seen in
2D neutrino-hydrodynamics simulations may be an artifact of the
artificial symmetry assumption and are actually nothing but 2D
convection in disguise. \citet{murphy_12} also noted that nonlinear
convection theory appears to explain the 3D flow properties of their
models without the need of invoking the SASI as an additional 
instability. At first glance, these findings seem to be in line 
with other 3D studies relying on a similar light-bulb methodology
\citep{iwakami_08,nordhaus_10,hanke_12,couch_12} or on a gray
neutrino transport approximation with chosen neutrino 
luminosities imposed at an inner grid boundary
\citep{wongwathanarat_10,wongwathanarat_12,mueller_e_12}. In contrast, \citet{mueller_12b}
demonstrated by self-consistent, 2D, general relativistic (GR) 
simulations with sophisticated transport that genuine SASI activity 
remains possible for sufficiently small shock stagnation radius 
(caused by high mass accretion rates
in the particular case of a $27 M_\odot$ progenitor of
\citealt{woosley_02}). This suggests that details of the conditions
may decide about the growth of the SASI, and that these conditions 
may not only depend on the properties of the progenitor star but also 
on the exact behavior of the stalled shock, which again depends on 
a reliable treatment of the neutrino physics. The mentioned 3D
models might simply have missed the sweet spot for SASI growth
in parameter space. 

Recent Newtonian simulations by \citet{takiwaki_12} and GR 
simulations by \citet{kuroda_12} and \citet{ott_12} are first,
tentative steps to higher sophistication in 3D models, but these
works were still focused on a few progenitors only and employed
crude neutrino transport methods with various simplifications
concerning the description of neutrino propagation, of 
neutrino interactions, and of the energy dependence of
the transport. \citet{ott_12}, using a neutrino leakage scheme
and studying the same $27 M_\odot$ progenitor as \citet{mueller_12b},
observed low-level SASI activity at early times, which was eventually
suppressed in models which exploded because of artificially 
enhanced neutrino heating. 
Their neutrino treatment, however, is not on par with the 
multi-group ray-by-ray-plus transport of \citet{mueller_12b} 
and therefore comparisons with the more sophisticated 2D models
of \citet{mueller_12b} should be made only with great caution 
and reservation.

In view of the poor exploration of conditions in collapsing
stellar cores in 3D so far and considering
the substantial approximations that have been made in 3D 
supernova simulations until now, the strong opinions uttered about 
the dominance of neutrino-driven convection \citep{burrows_12,murphy_12,
dolence_13} and the categorical
rejection of an important role of the SASI in ``realistic'' 3D
supernova models \citep{burrows_13} are disturbing. Actually, it is not
overly astonishing that the highly simplified setup investigated 
by these authors did not show signs of any pronounced SASI activity.
The growth of SASI modes was clearly disfavored in their models 
by several aspects.
Neglecting neutrino losses from the neutron star interior (above an
optical depth of about unity) and using a relatively stiff nuclear
equation of state the authors prevented the neutron star from 
shrinking below $\sim$60\,km. Correspondingly, the shock radius
remained rather large, in which case the postshock velocities were
relatively small and the advection timescale of the accretion flow 
through the gain layer was relatively long. This damped the
development of the SASI, whose growth rate is roughly proportional
to the inverse of the advection timescale (cf.\ \citealp{scheck_08}).
At the same time such conditions supported neutrino-driven convection,
which preferably develops in the accretion flow for ratios $\chi$ of 
the advection time to the local buoyancy timescale above a critical
value of $\chi \approx 3$ \citep{foglizzo_06}.

Although the theoretical understanding of the growth conditions of the
SASI and neutrino-driven convection is mostly based on linear theory
(e.g., \citealp{foglizzo_07,laming_07,yamasaki_07,yamasaki_08,guilet_12}), the
predictions were found to be consistent with the behavior seen in 2D
hydrodynamical simulations of accretion shocks in collapsing stellar
cores \citep{scheck_08} and in full-scale 2D supernova models
\citep{mueller_12b}. Similarly, although it is a priori not clear
whether the SASI sloshing motions \citep{blondin_03} and spiral modes
\citep{blondin_07,fernandez_10} observed in adiabatic accretion flows
in 2D and 3D, respectively, or in shallow-water experiments
\citep{foglizzo_12}, are conclusive for phenomena of relevance in the
convectively unstable environment of the neutrino-heated layer,
violent bipolar shock oscillations with SASI-typical characteristics
were also identified to determine the evolution of the stalled
supernova shock in some progenitors and preexplosion phases (e.g.,
\citealp{scheck_08,mueller_12b}).  While the phenomenon of the SASI as
an advective-acoustic instability is not generically linked to 2D,
strong SASI activity has so far not been detected in full-scale 3D
supernova simulations and it was speculated that its amplitude could
be reduced by the kinetic energy being shared between three instead of
two dimensions \citep{iwakami_08}, that the absence of a
flow-constraining symmetry axis might disfavor coherent mass motions
of low spherical harmonics modes \citep{burrows_13}, or that
neutrino-driven buoyancy is generally the fastest growing nonspherical
instability in supernova cores \citep{burrows_12}.

\begin{figure*}
\begin{center}
\includegraphics[width=.32\textwidth]{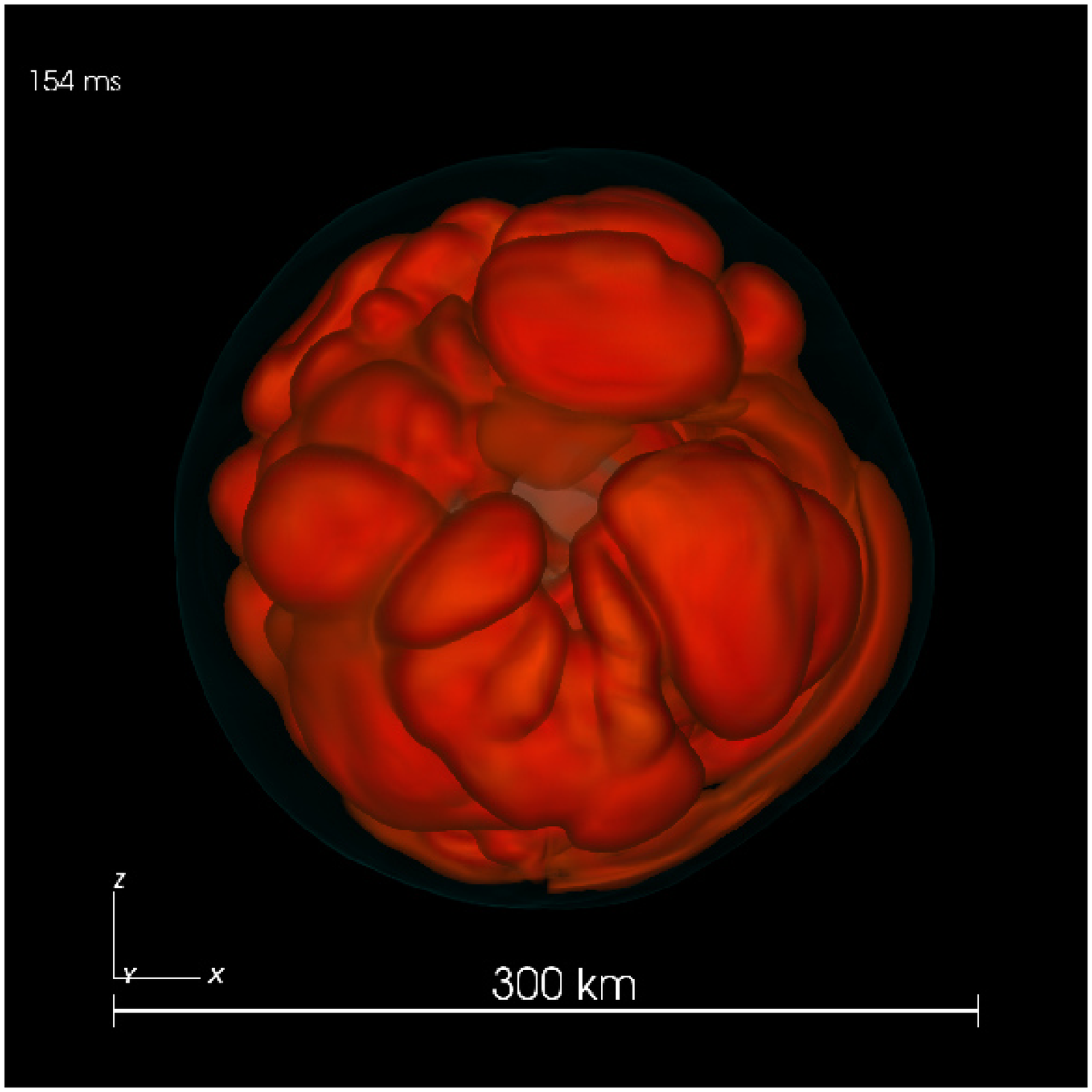}
\includegraphics[width=.32\textwidth]{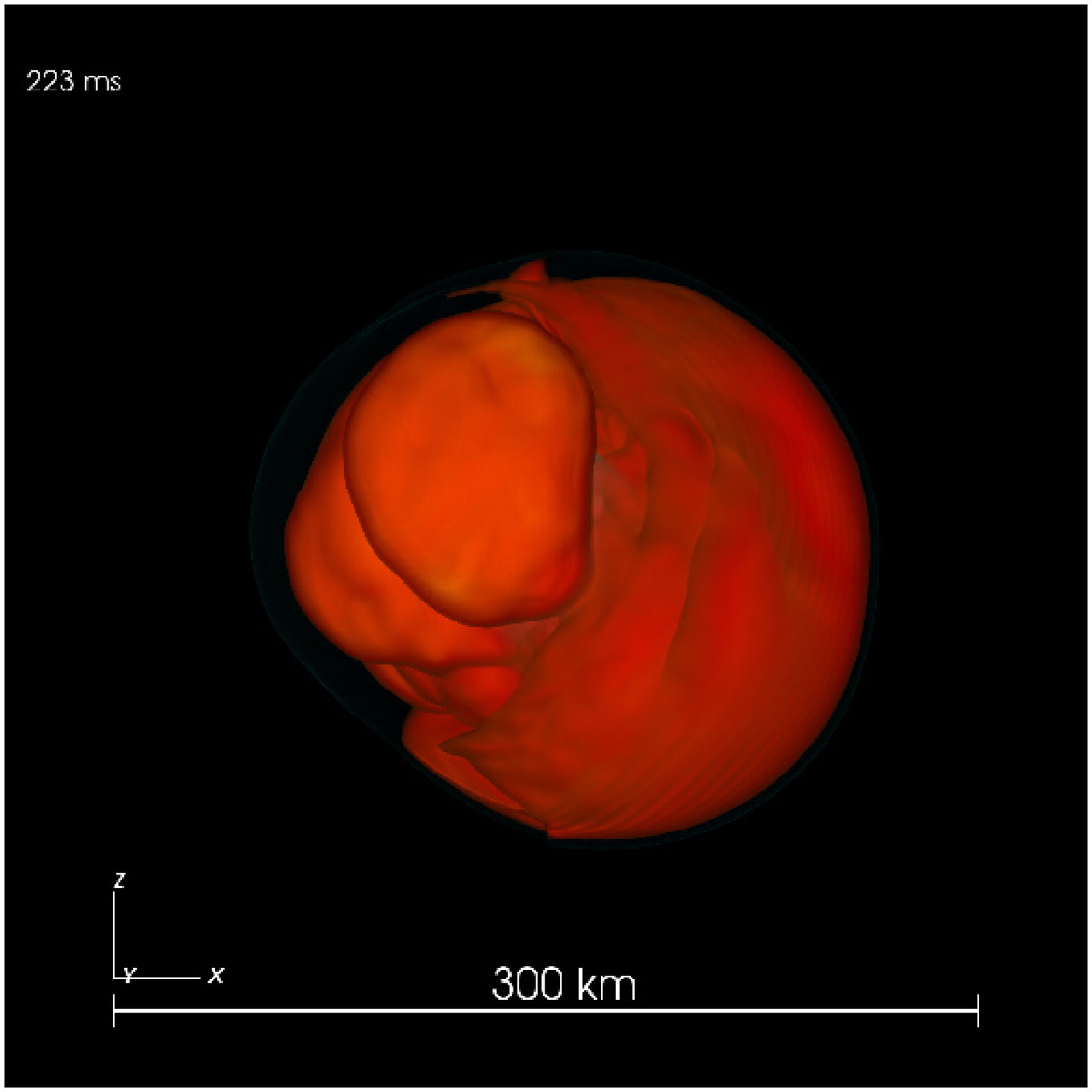}
\includegraphics[width=.32\textwidth]{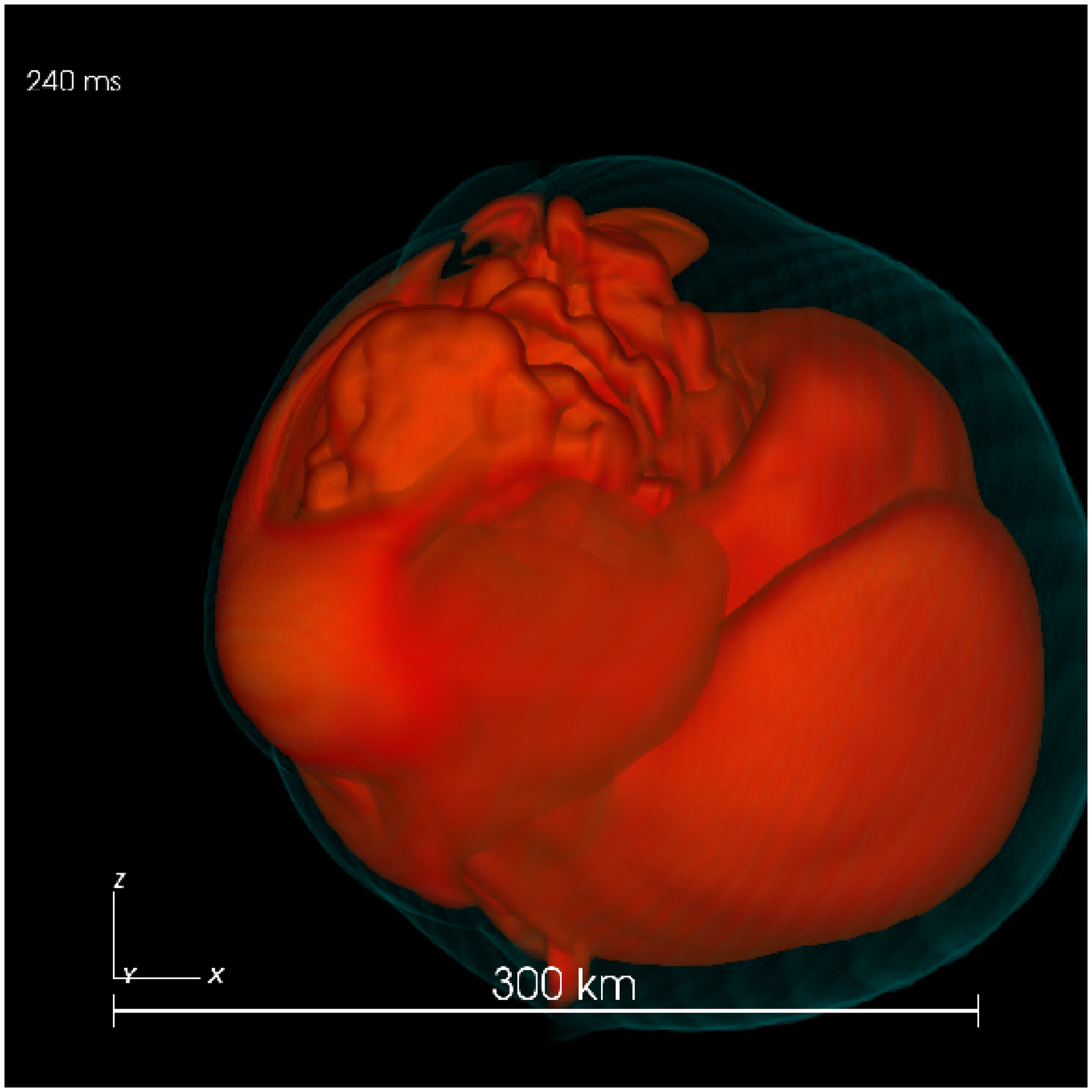}\\
\includegraphics[width=.32\textwidth]{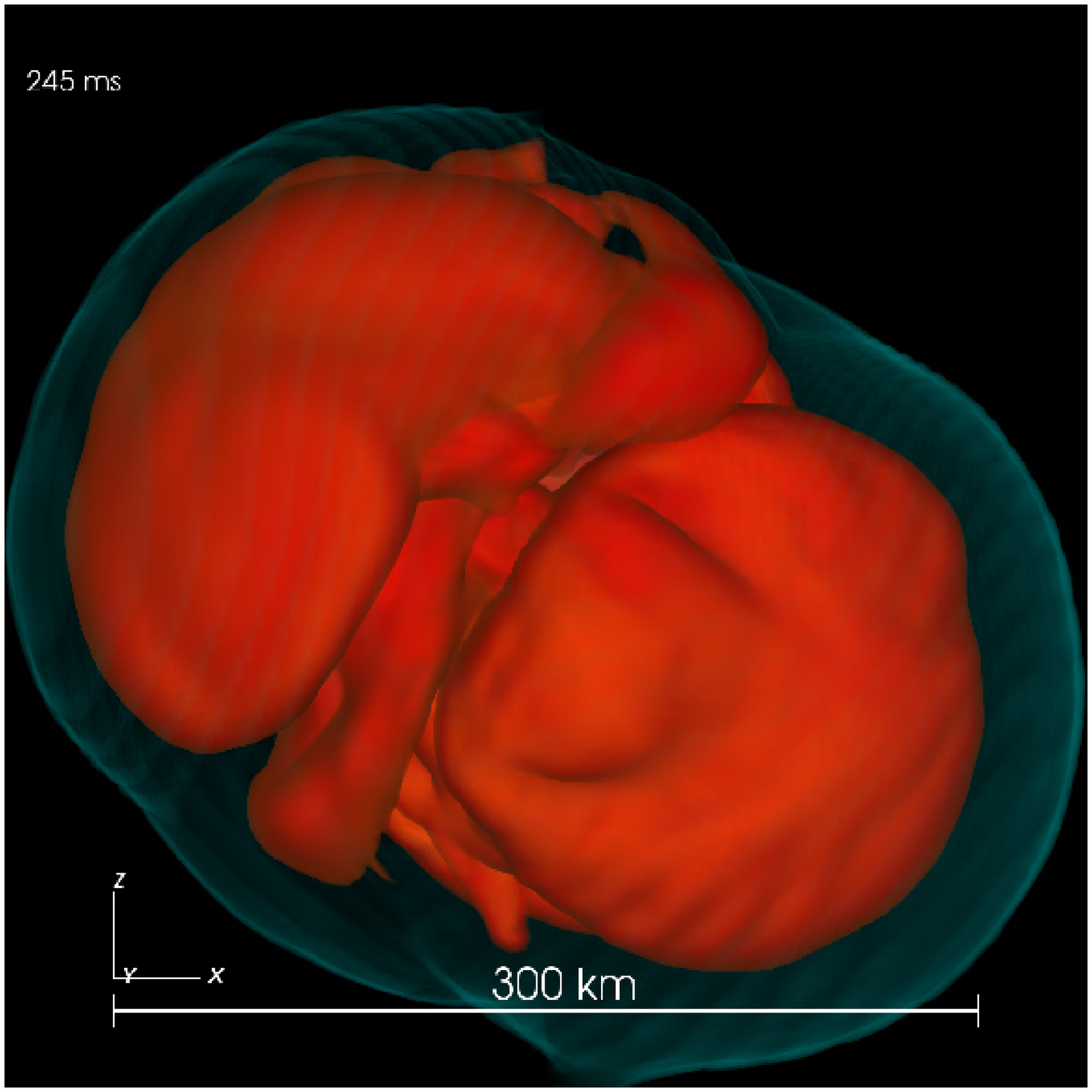}
\includegraphics[width=.32\textwidth]{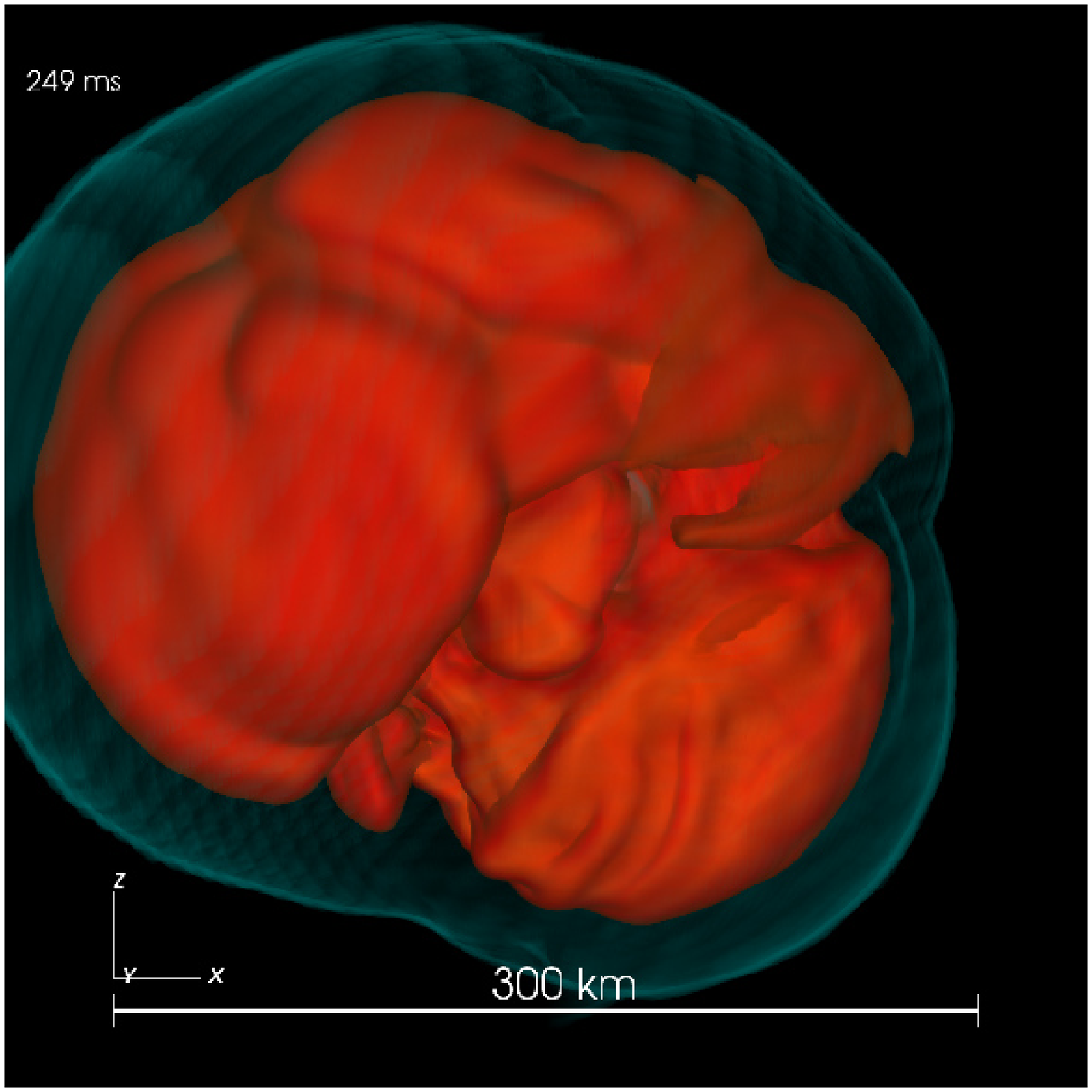}
\includegraphics[width=.32\textwidth]{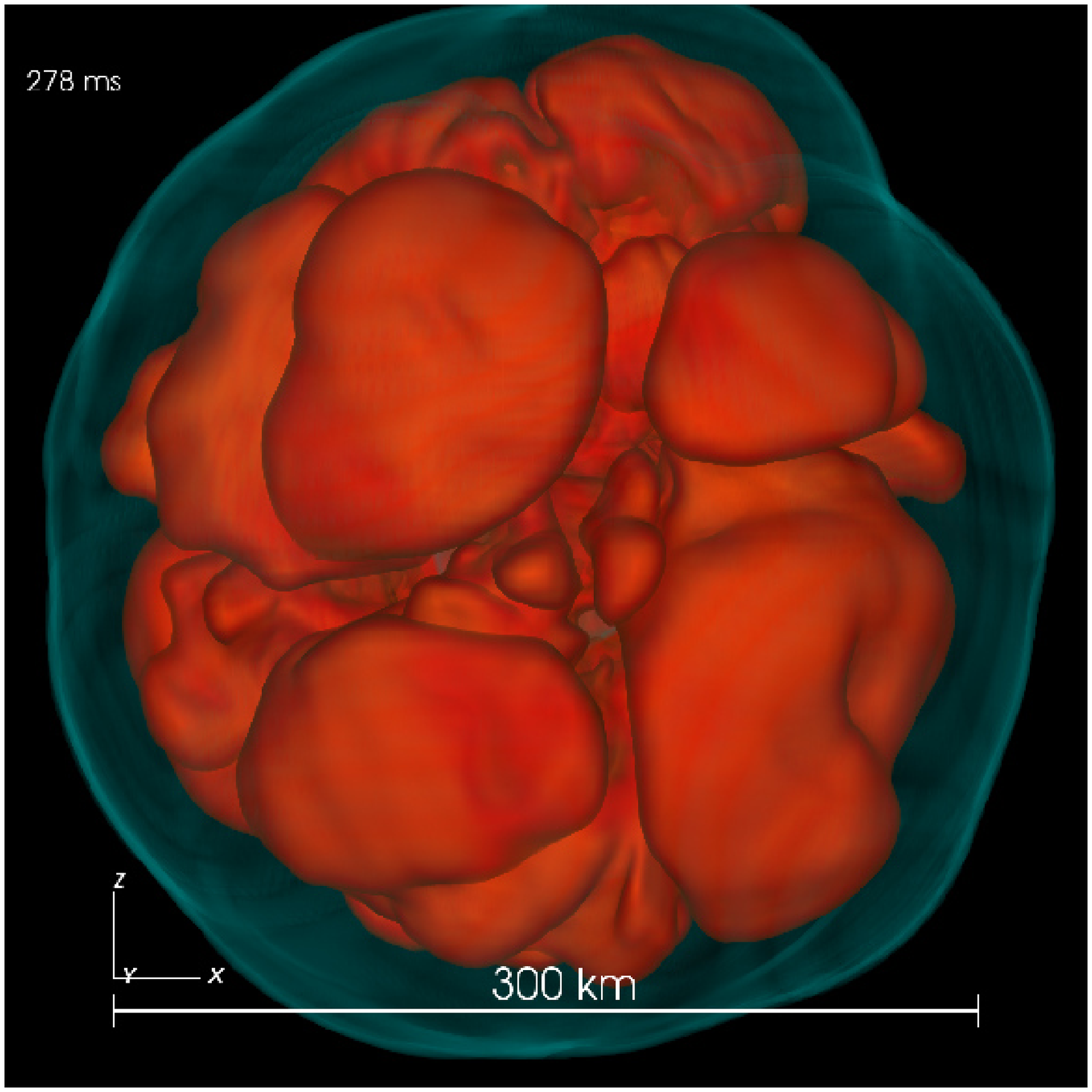}
\caption{\label{fig:3dsasi}
 Snapshots of phases with convective and SASI activity in the
 evolution of the 27\,$M_\odot$ model at 154\,ms, 223\,ms, 240\,ms ({\em upper
 panels, from left to right}), 245\,ms, 249\,ms, and 278\,ms ({\em lower panels,
 from left to right}). The volume rendering visualizes
 surfaces of constant entropy: The outer, bluish, semi-transparent surface
 is the supernova shock, the red surfaces are entropy structures in the
 postshock region. The upper left panel displays mushroom-like plumes
 of expanding, high-entropy matter that are typical of neutrino-driven
 buoyancy. The upper middle and right plots and the lower left and middle
 panels show distinctly different entropy
 structures of dipolar (and quadrupolar) asymmetry, which engulf the
 still visible buoyant plumes with their higher-order spherical harmonics mode
 pattern. The entropy asymmetries of $\ell = 1,\,2$ character are caused by
 global shock sloshing motions, which create hemispheric high-entropy shells 
 in phases of shock expansion. 
 At 223\,ms and 240\,ms the shock has pushed towards
 the lower right corner of the panels whereas at 245\,ms and 249\,ms it is in
 a phase of violent expansion motion towards the upper left corner of the
 plots. All stages exhibit a strong deformation of the shock. At 278\,ms the
 vivid SASI phase is over, the shock is more spherical again, and the
 postshock entropy structures correspond to neutrino-driven plumes.
  }
\end{center}
\end{figure*}

In this paper, we report about unambiguously identified SASI activity
in the first 3D simulation with detailed neutrino transport of the 27\,$M_\odot$ progenitor model 
that was investigated by \citet{mueller_12b}. For this 3D simulation
we employed the \textsc{Prometheus-Vertex} neutrino-hydrodynamics code
with detailed ray-by-ray-plus neutrino transport and the sophisticated
treatment of energy-dependent neutrino interactions also applied in
previous 2D simulations with this code (e.g., \citealp{buras_06_a,marek_09}) 
and in the 2D GR models of 
\citet{mueller_12,mueller_12b,mueller_13}. Contrary to the claims discussed above, our model shows
that despite the presence of neutrino-driven convection, the SASI can 
grow no less vigorously in 3D (without any coordinate grid-imposed 
symmetry) than in 2D as long as small shock radii are maintained
(in the 27\,$M_\odot$ star because of high mass accretion rates) and
guarantee favorable growth conditions. We also
observe the development of a clear spiral mode. SASI shock motions
appear to be diminished only when the accretion rate drops after
the Si/SiO shell interface reaches the shock and the shock is able
to expand to considerably larger radii. The variation of the relative
strengths of neutrino-driven convection and SASI sloshing is consistent
with experience and understanding based on previous 2D simulations.
In order to demonstrate this more directly we also performed parametric
2D and 3D core-collapse and explosion simulations by using simplified, 
gray neutrino transport as in \citet{mueller_e_12}
and \citet{wongwathanarat_10,wongwathanarat_12} to enhance the 
computational efficiency. Excising the high-density core and replacing it
by a contracting inner boundary condition, at which neutrino luminosities
with chosen values could be imposed, allowed us to control the supernova
core conditions. In these simulations we studied a 25\,$M_\odot$
progenitor and made use of an axis-free Yin-Yang grid
\citep{wongwathanarat_10b} instead of a polar grid.
These simulations did not only show that the discussed SASI 
phenomenon can play a role in different progenitor stars and is
not dependent on the choice of a particular computational grid, they
also confirmed that SASI mass motions in 3D can be triggered by the 
same conditions that are known to be favorable in 2D.

Our paper is structured as follows. In Sect.~\ref{sec:27msunmodel} we
discuss the results of our fully self-consistent neutrino-hydrodynamics
simulation of the 27\,$M_\odot$ progenitor, describe briefly the numerical methods
and modeling setup, analyze differences of the convective and SASI
activity between 2D and 3D simulations, and present evidence for the
temporary development of a spiral mode. In Sect.~\ref{sec:para25sunmodel}
we compare 2D and 3D results for three simulations of the 25\,$M_\odot$
model with our simplified modeling setup, whose main features we also 
summarize before we investigate the dependence of the development of
SASI activity on the parametrically regulated conditions in the
postshock and neutrino-heating layer. A discussion and conclusions 
will follow in Sect.~\ref{sec:conclusions}.

\begin{figure*}
\begin{center}
% \plotone{overview1.eps}
    \includegraphics[width=\textwidth]{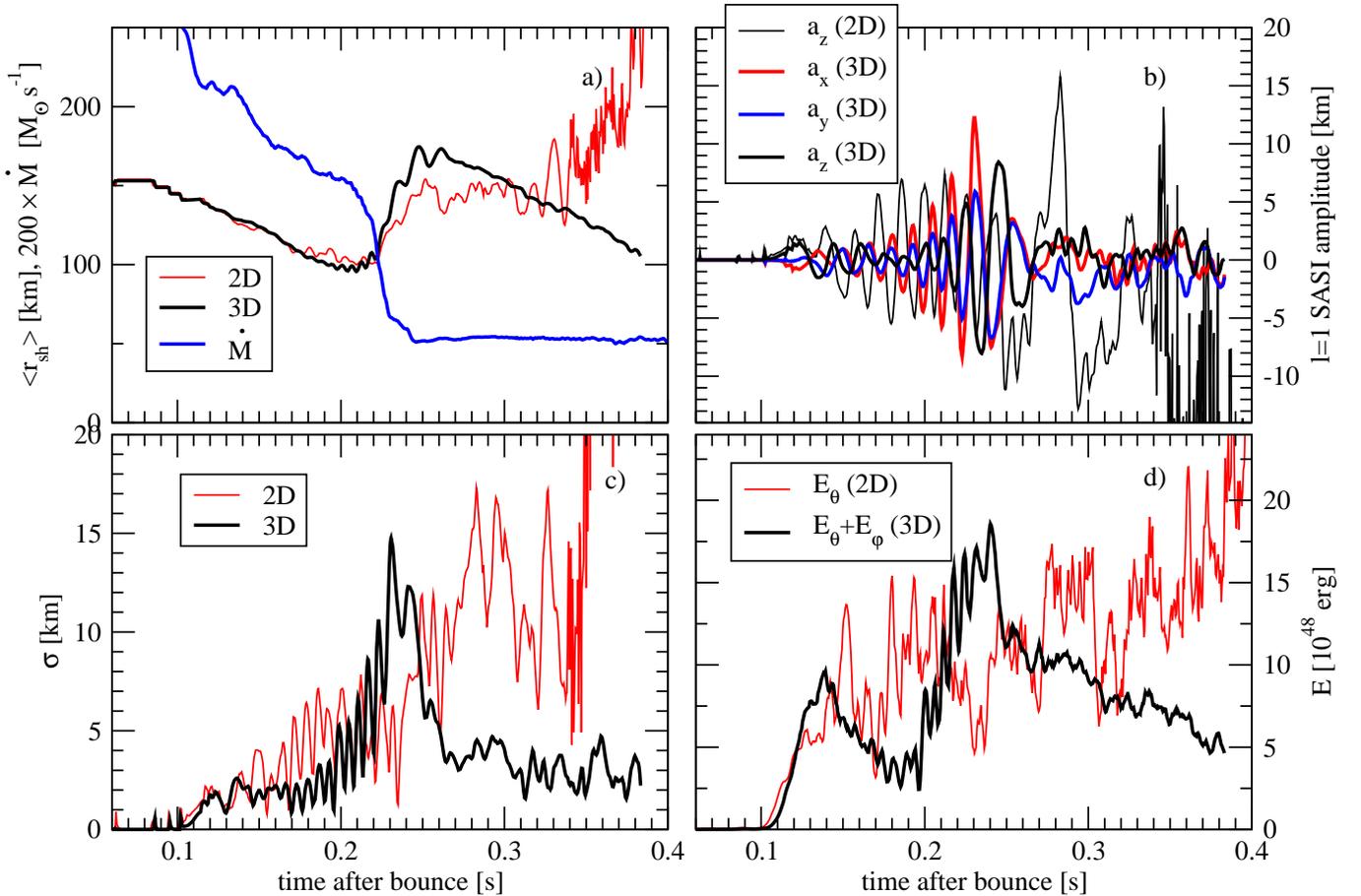}
  \caption{Comparison of SASI activity and shock evolution in the
2D and 3D simulations of the 27\,$M_\odot$ model.
{\em Panel a}:
average shock radius ($\langle r_\mathrm{sh} \rangle = a_0$) and
mass accretion rate of the collapsing stellar core at 400\,km;
{\em panel b}: components of the SASI $\ell=1$ amplitude vector;
{\em panel c}:
rms shock deformation $\sigma$; {\em panel d}: kinetic energy of non-radial
mass motions in the gain layer.
    \label{fig:overview}}
\end{center}
\end{figure*}

\section{Fully self-consistent 3D core-collapse simulation of a 
27\,$M_\odot$ star}
\label{sec:27msunmodel}

\subsection{Numerical methods and modeling setup}
\label{sec:setup1}
The calculations for the 27\,$M_\odot$ model in 2D and 3D were 
performed with the elaborate neutrino-hydrodynamics code 
\textsc{Prometheus-Vertex}. This supernova simulation tool
combines the hydrodynamics solver
\textsc{Prometheus} \citep{fryxell_89}, which is a dimensionally-split
implementation of the piecewise parabolic method (PPM) of
\citet{colella_84}, with the neutrino transport module \textsc{Vertex}
\citep{rampp_02}. \textsc{Vertex} solves the energy-dependent moment
equations for the neutrino energy and momentum density (with full
velocity dependence) for spherically symmetric transport problems 
defined to be associated with every angular bin of the polar grid 
(``radial rays'') used for the multi-dimensional simulations. The
moment equations are closed by a variable Eddington factor relation
that is provided by the formal solution of a model Boltzmann
equation. An up-to-date set of neutrino interaction rates is included
in \textsc{Vertex} (see, e.g., \citealt{mueller_12}). In the
multi-dimensional case, our ray-by-ray-plus approach 
\citep{buras_06_a} includes non-radial neutrino advection and
pressure terms in addition to the radial transport solves.
In the simulations presented here, we assume monopolar
gravity, but include general relativistic corrections
by means of an effective gravitational potential \citep{marek_06}.

We simulate the evolution of the $27 M_\odot$ progenitor of
\citet{woosley_02}, which was previously investigated by
\citet{mueller_12b} and \citet{ott_12}, both in 2D and in 3D, using
the high-density equation of state (EoS) of
\citet{lattimer_91} with a nuclear incompressibility of $K=220
\ \mathrm{MeV}$. The models are computed on a spherical polar
coordinate grid with an initial resolution of $n_r \times n_\theta
\times n_\varphi = 400 \times 88 \times 176$ (3D) and $n_r \times
n_\theta = 400 \times 88$ (2D) zones. Later, refinements of the radial
grid ensure adequate resolution in the PNS surface
region. The innermost $10 \ \mathrm{km}$ are computed in spherical
symmetry in both cases to avoid excessive time-step limitations.
Seed perturbations for aspherical instabilities are imposed
by hand $10 \ \mathrm{ms}$ after bounce by introducing random
perturbations of $0.1\%$ in density on the whole computational
grid.

\begin{figure}
\begin{center}
% \plotone{spectrum.eps}
  \includegraphics[width=\columnwidth]{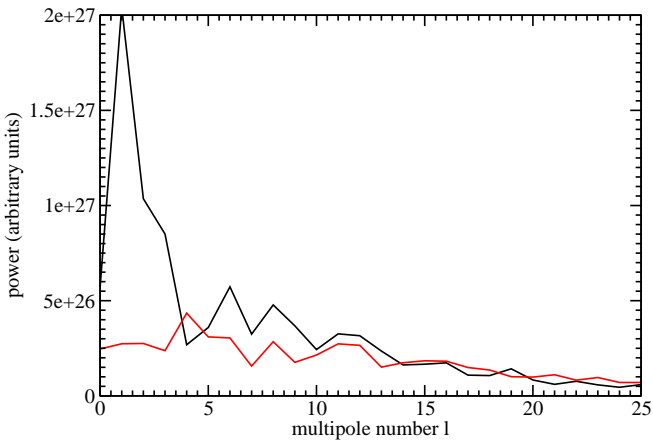}
  \caption{Power spectra of $v_\varphi$ of the 27\,$M_\odot$ 3D model
    sampled between $r=63$\,km and 80\,km at post-bounce
    times of 222\,ms (black) and 259\,ms (red).
    At 222\,ms, the strong SASI produces a distinctive peak
    at $l=1$, which is absent during the later, convection-dominated phase.
    \label{fig:spectrum}}
\end{center}
\end{figure}

\subsection{SASI activity: 2D versus 3D}
\label{sec:results}
While both the SASI and convection can lead to large-scale shock
deformations, the SASI is distinguished by a characteristic
oscillatory growth and in its nonlinear stage by the quasi-periodic,
oscillatory nature of the shock motions. In 2D, the artificial
symmetry constraint and the excitation of large-scale modes by the
inverse turbulent cascade could still produce a quasi-periodic
sloshing motion even in convectively-dominated models
\citep{burrows_12}, and a more refined analysis is necessary to
identify the SASI \citep{mueller_12b}.  However, in 3D the distinction
is much clearer, since large-scale shock deformations caused by
buoyancy-driven convection initially evolve randomly without any
identifiable periodicity and then grow \emph{monotonically} once they
reach a certain threshold amplitude \citep{burrows_12}. Periodic SASI
oscillations and large-scale shock deformations caused by convection
can therefore hardly be mistaken for each other in 3D.

Images of the entropy distribution in the postshock layer
(Figure~\ref{fig:3dsasi}) and in particular the corresponding movie of
our 3D simulation of the 27\,$M_\odot$ progenitor indeed provide a
clear hint that both distinctly different instabilities are at work in
the shocked accretion flow around the nascent neutron star.  The
instabilities develop nearly at the same time and are present
simultaneously for an extended period of the simulated postbounce
evolution. The first small mushroom-like Rayleigh-Taylor fingers of
neutrino-driven convection become visible around 80--100\,ms after
bounce to subsequently grow stronger and larger in angular size over a
timescale of some ten milliseconds. At about 125\,ms p.b.\ the rising
plumes begin to cause shock deformation and a modest amount of global
asphericity of the accretion layer. Until $\sim$155\,ms the activity
in the postshock layer is clearly dominated by neutrino-driven
buoyancy (Figure~\ref{fig:3dsasi}, upper left panel), but at $t\gtrsim
155$\,ms, during a phase of accelerated shock recession, coherent
entropy structures show up first. The corresponding low-mode spherical
harmonics pattern clearly differs from the buoyant mushrooms on
smaller angular scales. This phenomenon is associated with shock
oscillations, which quickly amplify to bipolar shock sloshing motions
and create characteristic, hemispheric high-entropy shells during
phases of fast shock expansion. These half-shells of shock-heated
matter engulf the buoyant bubbles of neutrino-driven convection in
deeper regions (Figure~\ref{fig:3dsasi}, upper middle and right and
lower left and middle panels). While the sloshing axis initially
wanders, it becomes more stable as the SASI sloshing of the shock
further grows in amplitude and violence between $\sim$195\,ms and
$\sim$240\,ms.  As a consequence, an expansion of the average shock
radius is driven even before the Si/SiO composition-shell interface
arrives at the shock and the mass accretion rate starts to drop
steeply at $t \sim 220$\,ms (Figure~\ref{fig:overview}, upper left
panel). The decrease of the accretion rate supports the shock
expansion, in course of which the bipolar, quasi-periodic shock
pulsations gain even more power. At $t\sim 225$\,ms a spiral mode
seems to set in for several revolutions before the average shock
radius reaches its maximum extension at $\sim 250$\,ms and the SASI
sloshing dies off at $t \gtrsim 260$\,ms. The presence of
large-amplitude spiral motions is reflected by considerable variations
of the mean shock radius between $230$\,ms and $260$\,ms. These
disappear when the SASI activity ceases at $t \sim 260$\,ms
(Figure~\ref{fig:overview}, upper left panel). Later on, until the end
of our 3D simulation, aspherical mass motions in the postshock layer
are dominated again by the buoyant plumes typical of neutrino-driven
convection (Figure~\ref{fig:3dsasi}, lower right panel).

This verbal description of the dynamical evolution of the postshock
accretion layer is supported by a detailed analysis based on 
several time-dependent parameters that quantify the characteristic
features of SASI activity. To this end we perform a time-dependent
decomposition of the angle-dependent shock position $r_\mathrm{sh}
(\theta,\varphi)$ into spherical harmonics $Y_\ell^m$:
\begin{equation}
  \label{eq:ylm_decomposition}
a_{\ell}^{m} = \frac{(-1)^{|m|}}{\sqrt{4\pi \left(2 \ell + 1\right)}}
\int_\Omega r_\mathrm{sh} (\theta,\varphi) Y_\ell^m (\theta,\varphi)\ud
\Omega \, .
\end{equation}
Here the $Y_\ell^m$ are real spherical harmonics with the same
normalization as used by \citet{burrows_12} and \citet{ott_12}.  With
this choice of basis functions, the coefficients with $\ell=1$ give the
angle-averaged Cartesian coordinates of the shock surface,
\begin{equation}
a_{1}^{-1}=\left \langle y_\mathrm{sh} \right \rangle =: a_y, \quad
a_{1}^{0}=\left \langle z_\mathrm{sh} \right \rangle =: a_z, \quad
a_{1}^{1}=\left \langle x_\mathrm{sh} \right \rangle =: a_x,
\end{equation}
and $a_0^0$ is just the average shock-radius $\langle r_\mathrm{sh} \rangle$.

The time evolution of the coefficients $a_x$, $a_y$ (3D), and $a_z$
(3D and 2D) is shown in panel~b of Figure~\ref{fig:overview}.  Both
in 2D and in 3D, the shock surface clearly oscillates in a
quasi-periodic manner until $\sim$260\,ms after bounce,
i.e.\ until shortly after the Si/SiO shell interface has reached the 
shock and the accretion rate has dropped considerably between $\sim$220\,ms
and 240\,ms p.b. The lower accretion rate
results in a pronounced expansion of the average shock radius (panel~a
of Figure~\ref{fig:overview}), which initially is stronger in
3D. However, the 2D model maintains large (albeit less regular) shock
oscillations, with the average shock radius eventually overtaking the
3D model at $\sim$300\,ms when an explosion develops
(i.e.\ somewhat later than in the GR simulation of
\citealt{mueller_12b})\footnote{It is not clear whether this difference
or how much of this difference is caused by GR effects, because the 
models in the present paper were simulated with a slightly different
treatment of the low-density equation of state, which led to a 
significant delay ($\sim$35\,ms) of the infall of 
the silicon layer and a correspondingly later arrival of the 
Si/SiO shell interface at the shock.}.
By contrast, the shock continues to recede in
the 3D run. The more optimistic evolution of the 2D model compared to
the failing 3D model at late stages is consistent with the findings of
\citet{hanke_12}, and could be due to the action of the inverse
turbulent energy cascade, which continues to feed energy into
large-scale modes in 2D.

\begin{figure}
\begin{center}
% \plotone{chi27.eps}
  \includegraphics[width=\columnwidth]{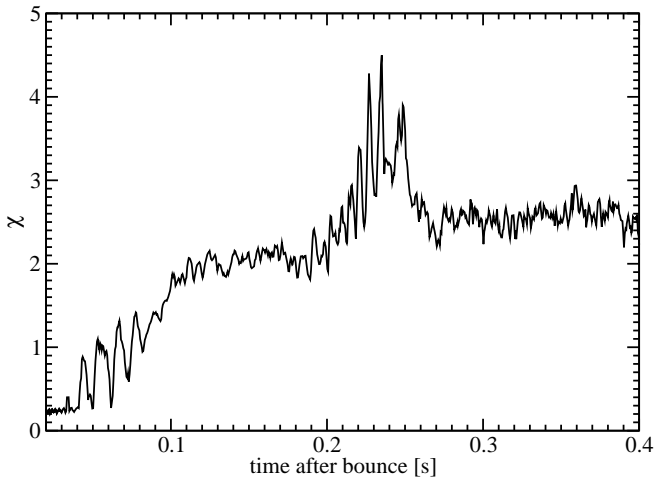}
  \caption{Evolution of the stability parameter $\chi$ for the gain layer
   of the 3D simulation of the 27\,$M_\odot$ progenitor. During
   most of the time $\chi < 3$. This suggests conditions in the
   postshock accretion flow which disfavor the growth of neutrino-driven
   convection relative to the development of the SASI in analogy to
   the 2D model discussed by \citet{mueller_12b}.
   \label{fig:chi27}}
\end{center}
\end{figure}

\begin{figure}
\begin{center}
% \plotone{convection_152ms.eps}
  \includegraphics[width=\columnwidth]{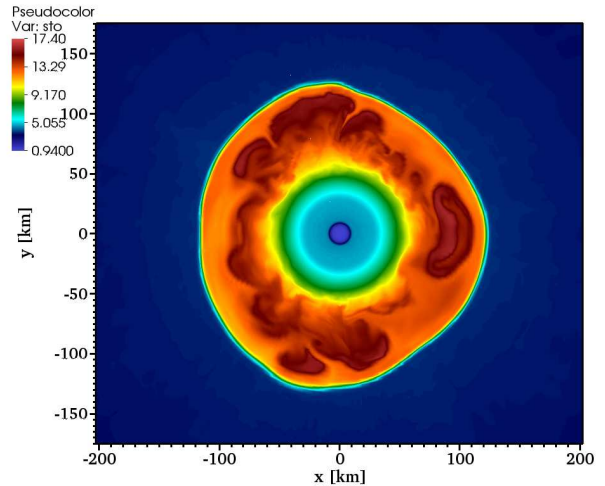}
  \caption{Snapshot of the entropy (color coded according to color
   bar at the upper left corner, in units of Boltzmann's constant
   $k_\mathrm{b}$ per nucleon) in the plane through the origin normal
   to the vector $\mathbf{n}=(-0.35, 0.93, 0.12)$ at a post-bounce
   time of 152\,ms in the 27\,$M_\odot$ 3D model. The high-entropy
   plumes with high-order spherical harmonics pattern suggest 
   buoyancy-driven convective overturn of neutrino-heated matter.
   \label{fig:convection}}
\end{center}
\end{figure}

However, the evolution of the two simulations prior to the
infall of the Si/SiO interface is remarkable: While the amplitude of
the $\ell=1$ mode is initially larger in 2D, the individual components
$a_x$, $a_y$, and $a_z$ of the $\ell=1$ amplitude vector in 3D
become comparable to $a_z$ in 2D around $200 \ \mathrm{ms}$, and
$a_x$ even reaches considerably bigger values. During this phase,
the SASI is undoubtedly \emph{stronger} in 3D than in 2D. Further
confirmation of this assessment is provided by the root-mean-square 
deviation
$\sigma(r_\mathrm{sh})$ of the shock radius from its average value 
(panel~c of Figure~\ref{fig:overview}):
\begin{equation}
  \sigma = 
\sqrt{
(4 \pi)^{-1} \int \left(r_\mathrm{sh}(\theta,\varphi) - 
\langle r_\mathrm{sh} \rangle  \right)^2 \, \ud \Omega
}
\end{equation}
For reasonably small amplitudes, $\sigma$ is also a measure for the 
total power of SASI amplitudes with different $\ell$:
\begin{equation}
\sigma \approx \sqrt{\sum_{\ell=1}^\infty \sum_{m=-\ell}^{\ell} 
|a_{\ell}^m|^2} \, .
\end{equation}
The same picture emerges when we consider the kinetic energies
$E_\theta$ and $E_\varphi$ associated with motions in the $\theta$- 
and $\varphi$-directions in the gain region,
\begin{equation}
E_\theta=\frac{1}{2}\int_{V_\mathrm{gain}} \rho v_\theta^2 \, \ud V, \quad
E_\varphi=\frac{1}{2}\int_{V_\mathrm{gain}} \rho v_\varphi^2 \, \ud V.
\end{equation}
As shown in panel~d of Figure~\ref{fig:overview}, the total
energy contained in non-radial motions is also larger in 3D during the
relevant phase around $\sim$230\,ms. In the period of continuous increase
of the SASI amplitude in the 3D model between $t\sim 155$\,ms
and $\sim 240$\,ms, the kinetic energy grows and $\sigma$ exhibits 
quasi-periodic modulations signaling the shock sloshing motions. 
Interestingly, during the phase of strongest SASI activity we
find rough equipartition between the kinetic energies of non-radial
motions $E_\theta + E_\varphi$, and the energy $E_r$ contained in
fluctuating radial velocities,
\begin{equation}
E_r=\frac{1}{2}\int_{V_\mathrm{gain}} \rho \left(v_r-\langle v_r \rangle \right)^2 \, \ud V,
\end{equation}
where $\langle v_r \rangle$ is the angle-averaged radial velocity.
This equipartition is apparently not a unique feature of 
buoyancy-driven turbulence (cf.\ \citealt{murphy_12}),
at least not as far as these volume-integrated
quantities are concerned.

A clear difference between SASI dominated and convection dominated
phases of the 27\,$M_\odot$ 3D model can be observed in the power 
spectrum of the azimuthal velocity $v_\varphi$ as a function of 
multipole order $\ell$. Figure~\ref{fig:spectrum} shows the spectra
during a SASI active phase (222\,ms p.b.) compared to the later time 
(259\,ms) when the SASI motions cease and convective plumes with
their higher-order multipole pattern determine the asphericities 
in the postshock
region again. The power spectra are evaluated with equation~(8) of
\citet{hanke_12} for $v_\varphi$ (weighted with the 
square root of the density) integrated over a radial region 
between 63\,km and 80\,km. The presence of the
SASI low-mode deformation at 222\,ms leads to a prominent peak
of the power spectrum at low multipole orders, which is absent 
in the later spectrum.

The shock deformation and SASI amplitude moderately increase as
the shock retreats between $\sim$100\,ms and 200\,ms p.b. (The same
is true for the specific nonradial kinetic energy in the gain layer,
although the kinetic energy decreases temporarily because of the
decreasing mass in the gain region.)
It is noteworthy that the few tens of milliseconds of stronger SASI
activity in 3D coincide with a phase of more rapid shock
expansion than in 2D.
Conceivably, the more energetic SASI motions provide a
stronger push against the pre-shock ram pressure. A close
examination of the $\langle r_\mathrm{sh} \rangle$ in
Figure~\ref{fig:overview} shows that some (oscillatorily
modulated) SASI-aided shock expansion seems to set in around
$190 \ \mathrm{ms}$, i.e.\ already before the rapid drop of the
preshock mass-accretion rate that begins at $\sim$220\,ms.
Therefore one might speculate that with slightly more
time available for the growth of the SASI, the extra support
by nonradial SASI motions might have driven the 3D
model over the threshold for a neutrino-powered runaway
expansion of the shock after the infall of the Si/SiO interface.

%\begin{figure*}
\begin{figure}
\begin{center}
% \plottwo{sasi_vector1.eps}{sasi_vector2.eps}
%  \includegraphics[width=.45\textwidth]{f6a.eps}\hspace{10pt}
%  \includegraphics[width=.45\textwidth]{f6b.eps}
\includegraphics[width=.45\textwidth]{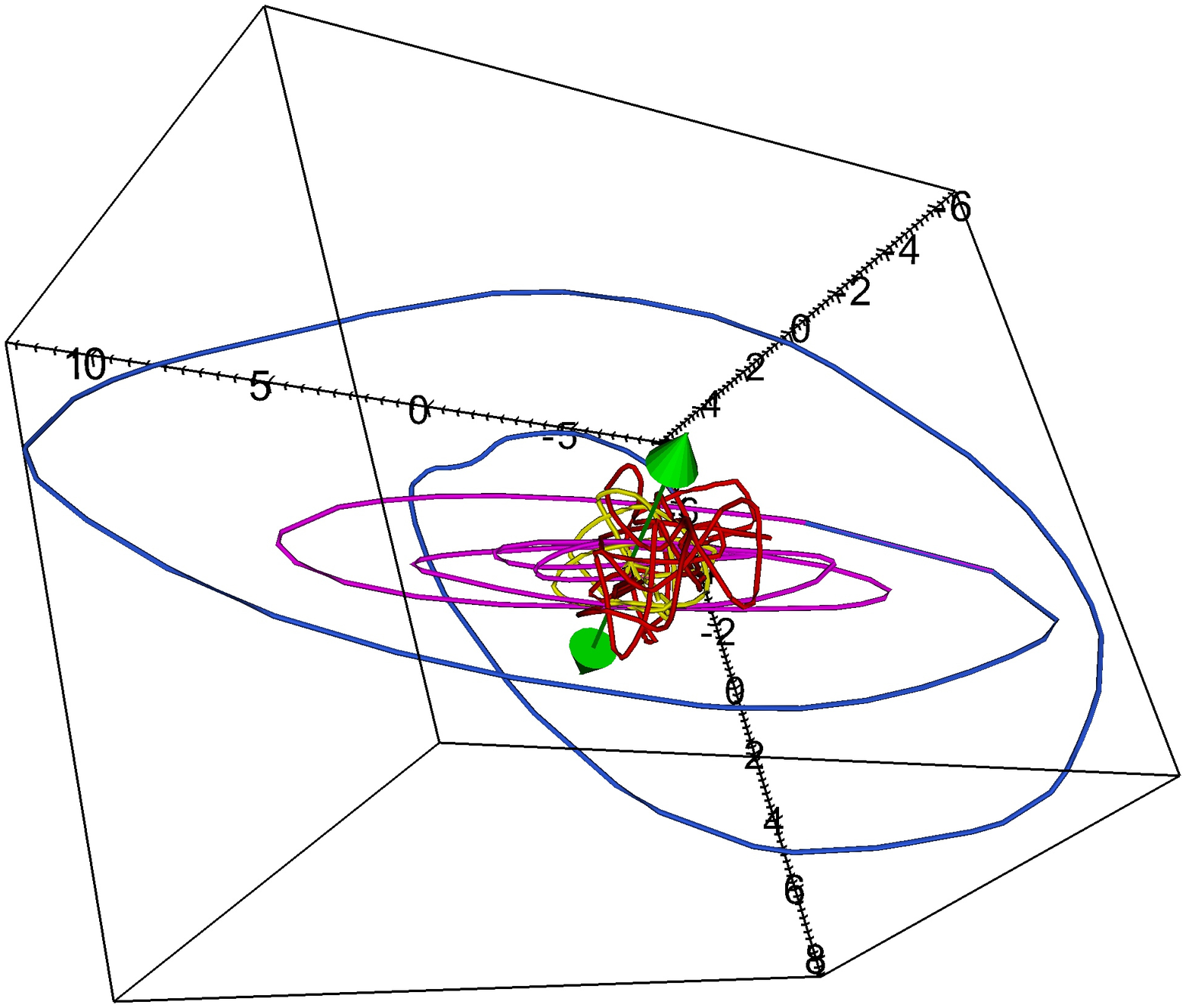}\\
\includegraphics[width=.45\textwidth]{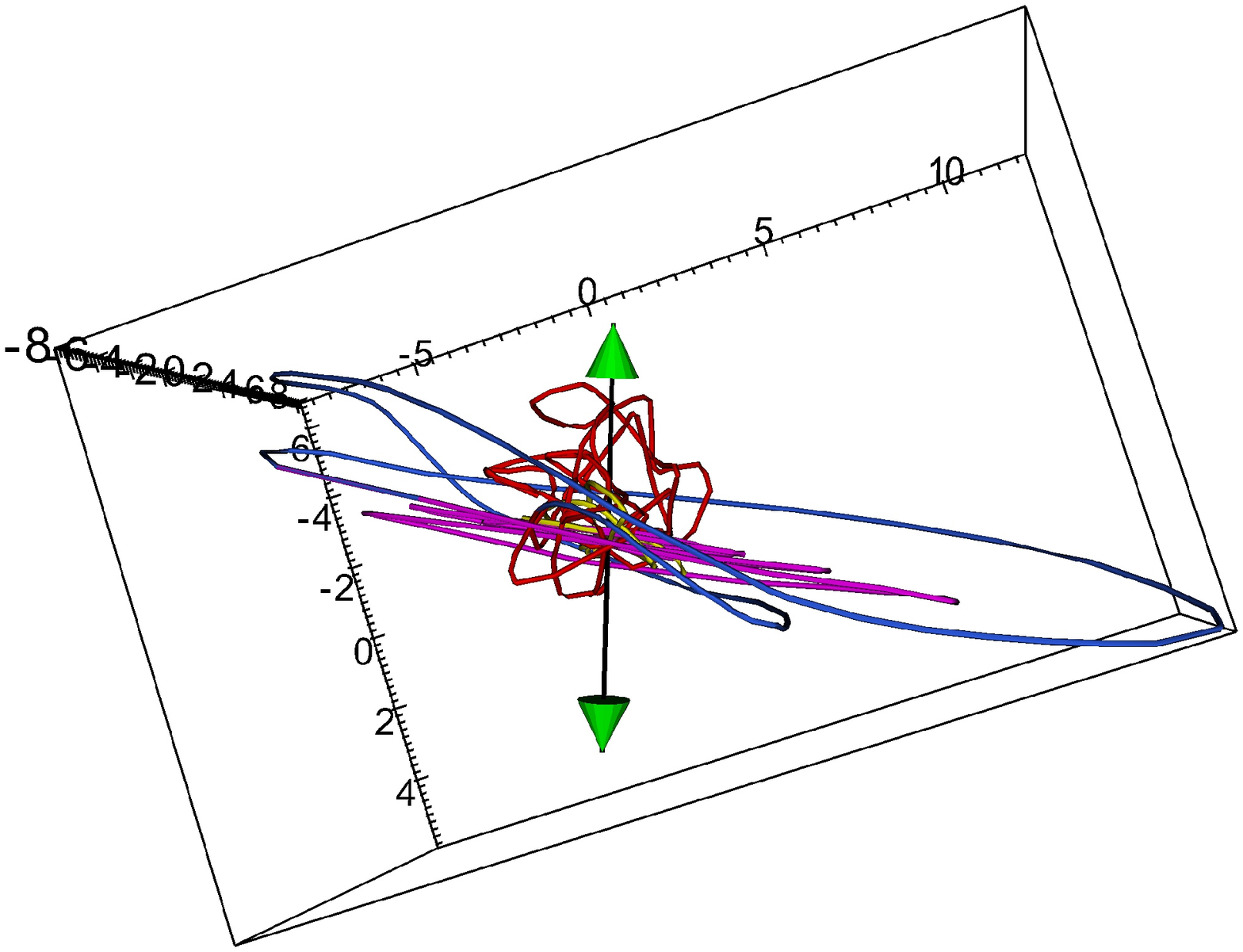}\\
\includegraphics[width=.45\textwidth]{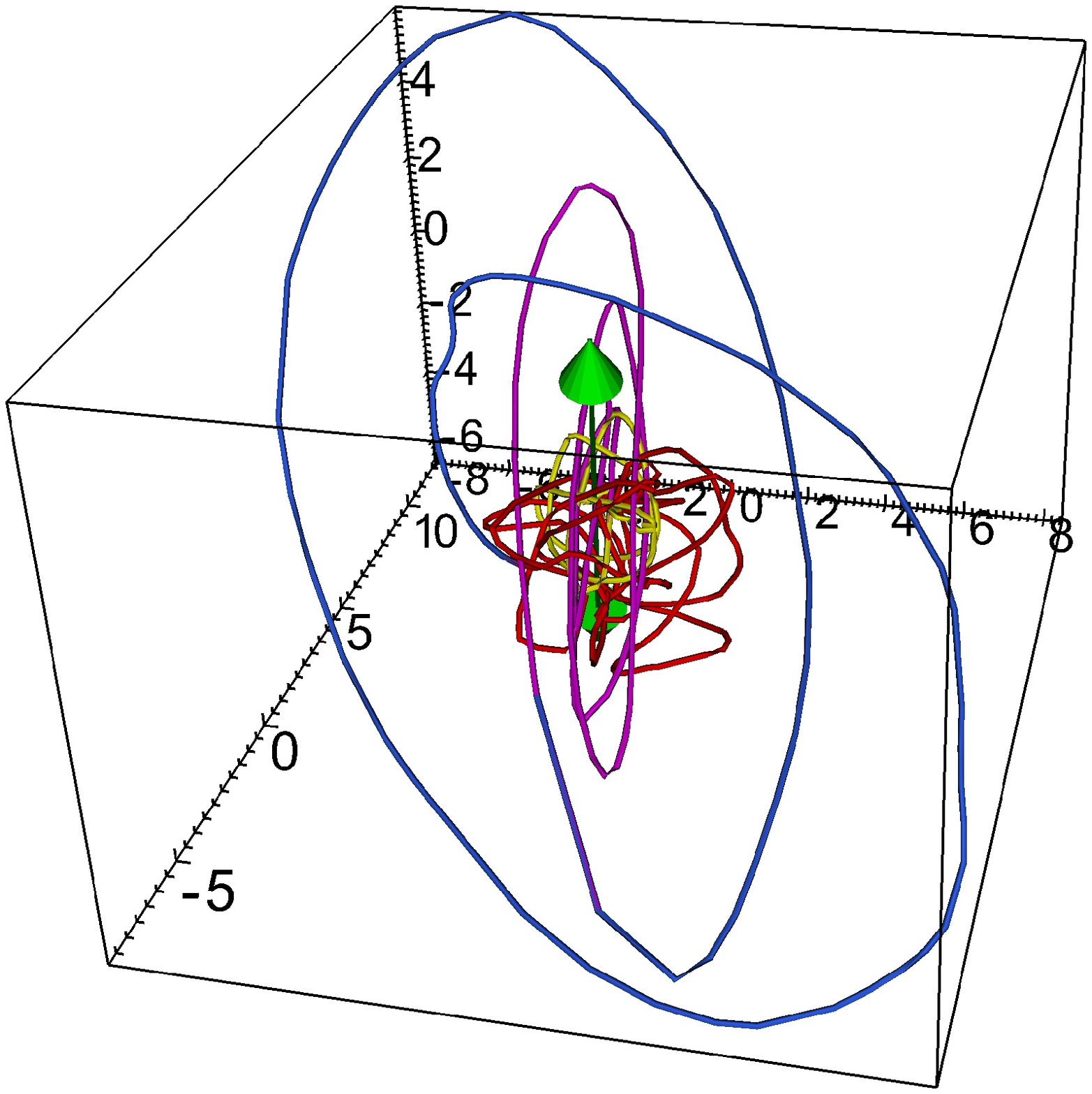}
  \caption{Evolution of the $\ell=1$ amplitude vector $\mathbf{a}_1$
for the 27\,$M_\odot$ simulation from three different viewing angles.
Different colors are used for the
phase up to 177\,ms (yellow), the phase of strong 
SASI sloshing activity (magenta, up to 225\,ms), clearly developed
SASI spiral motion (blue, up to 265\,ms), and the late, SASI-quiet
phase (red). The arrows indicate the vector $\mathbf{n}=(-0.35, 0.93, 0.12)$
and its counter-vector perpendicular to the rotational plane of the SASI 
(see also Figure~\ref{fig:spiral}). Note that the transition from SASI
sloshing to spiral behavior is gradual (and associated with a strong
growth of the angular momentum in the gain layer; 
Figure~\ref{fig:angular_momentum}), and the color coding is based
on eye inspection rather than a precise definition.
    \label{fig:3d_amplitude}}
\end{center}
%\end{figure*}
\end{figure}

\begin{figure*}
\begin{center}
% \plottwo{spiral_223ms_new.eps}{spiral_227ms_new.eps}
% \plottwo{spiral_231ms_new.eps}{spiral_235ms_new.eps}
  \includegraphics[width=.45\textwidth]{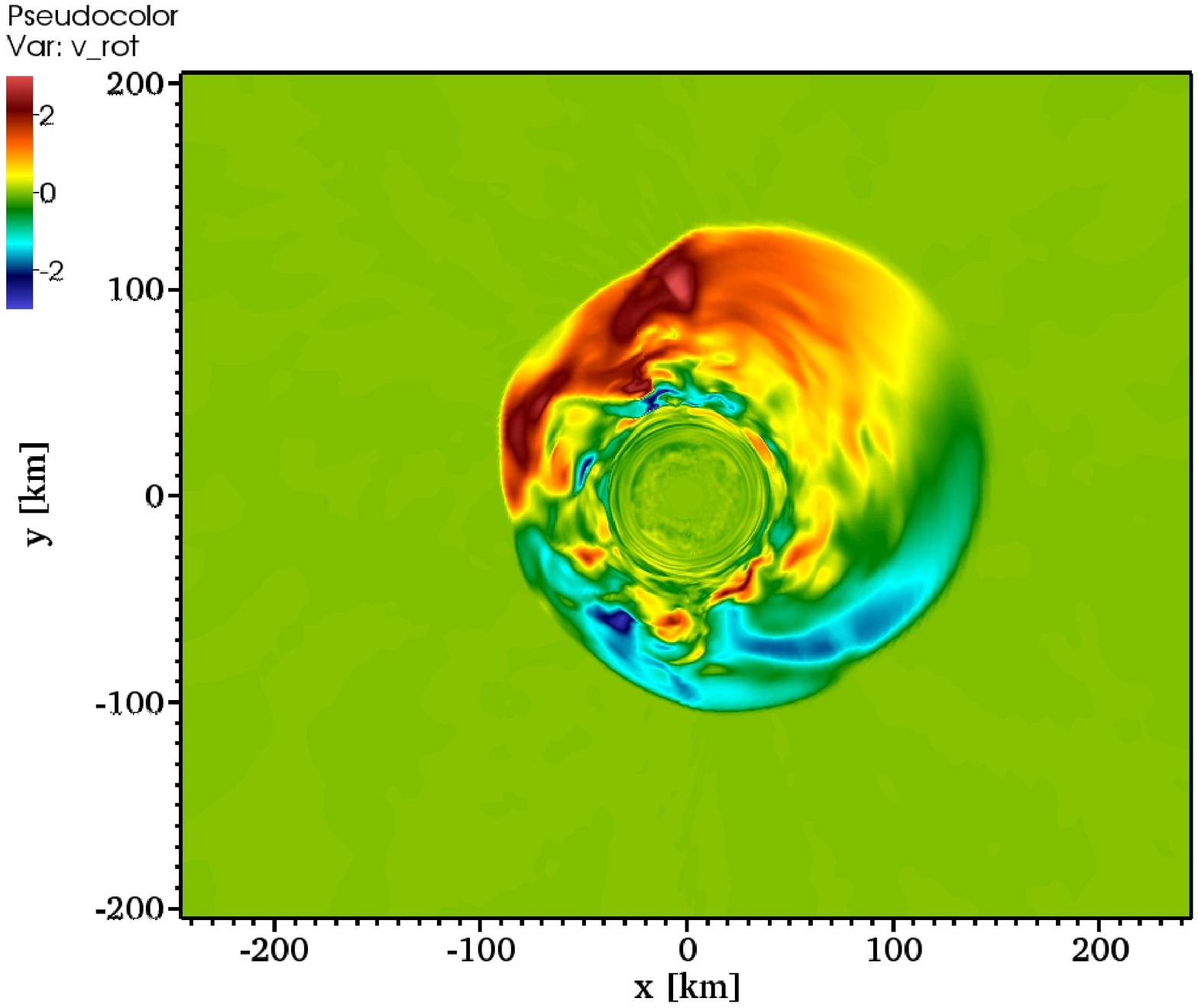}\hspace{8pt}
  \includegraphics[width=.45\textwidth]{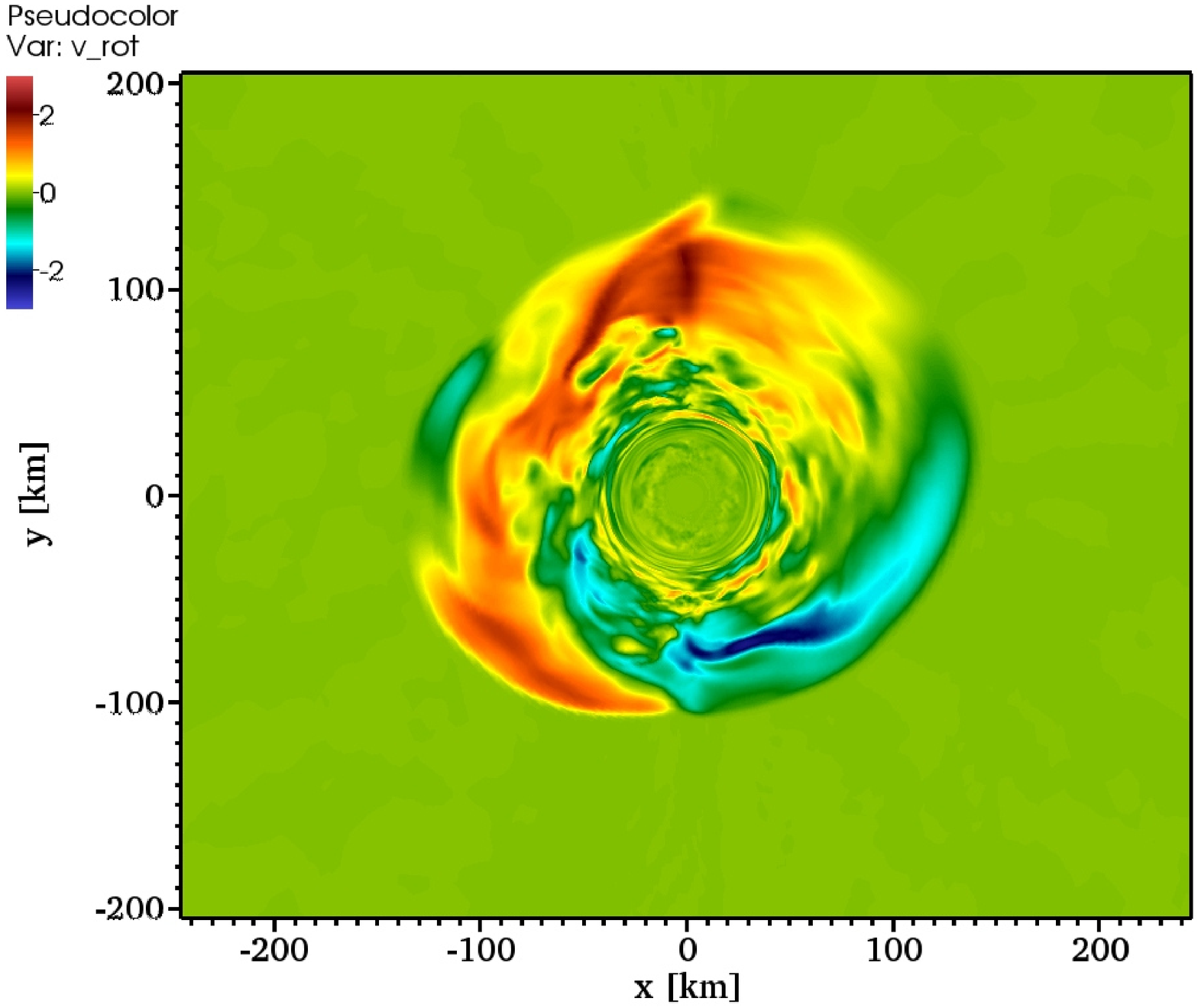}\\
  \includegraphics[width=.45\textwidth]{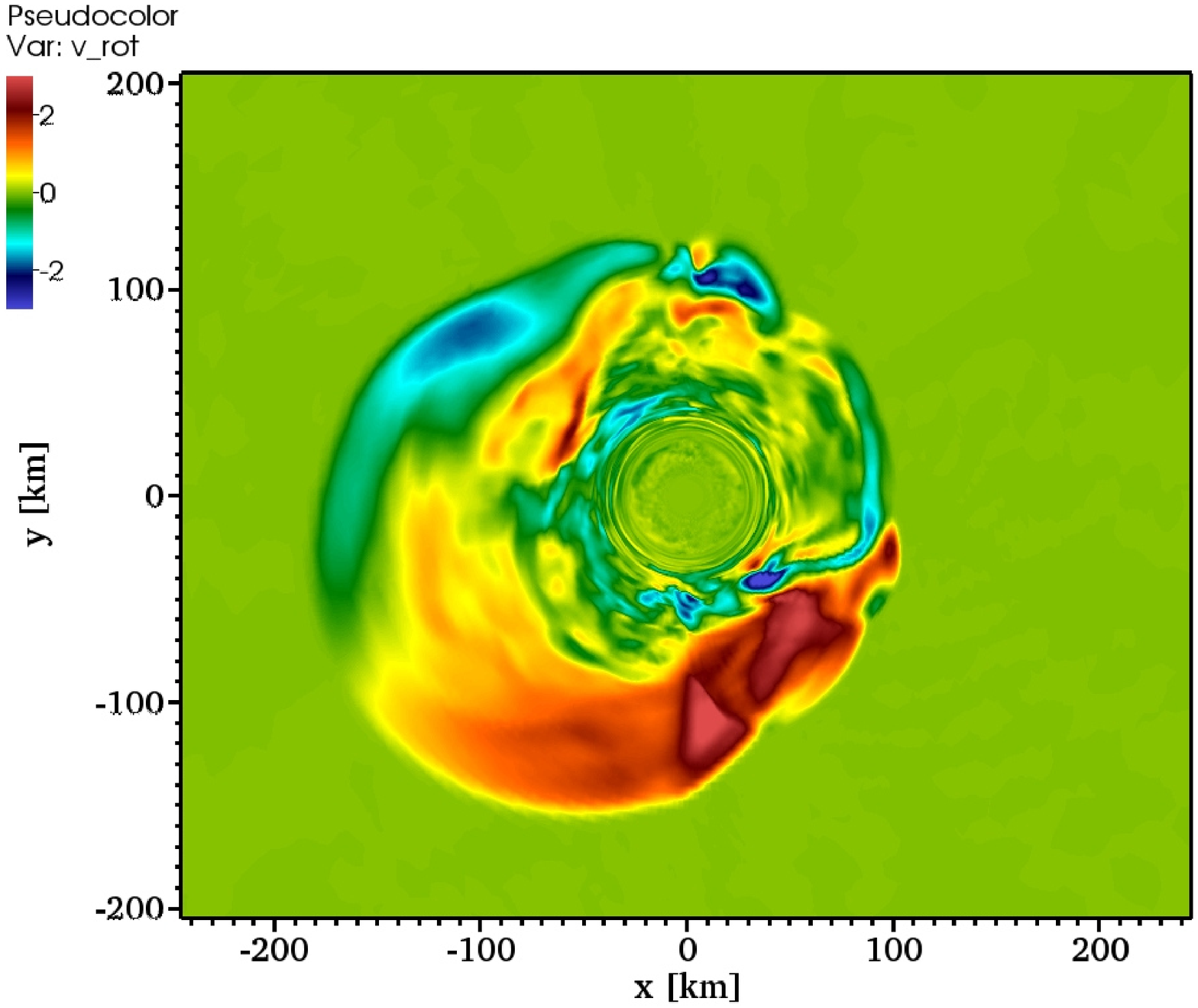}\hspace{8pt}
  \includegraphics[width=.45\textwidth]{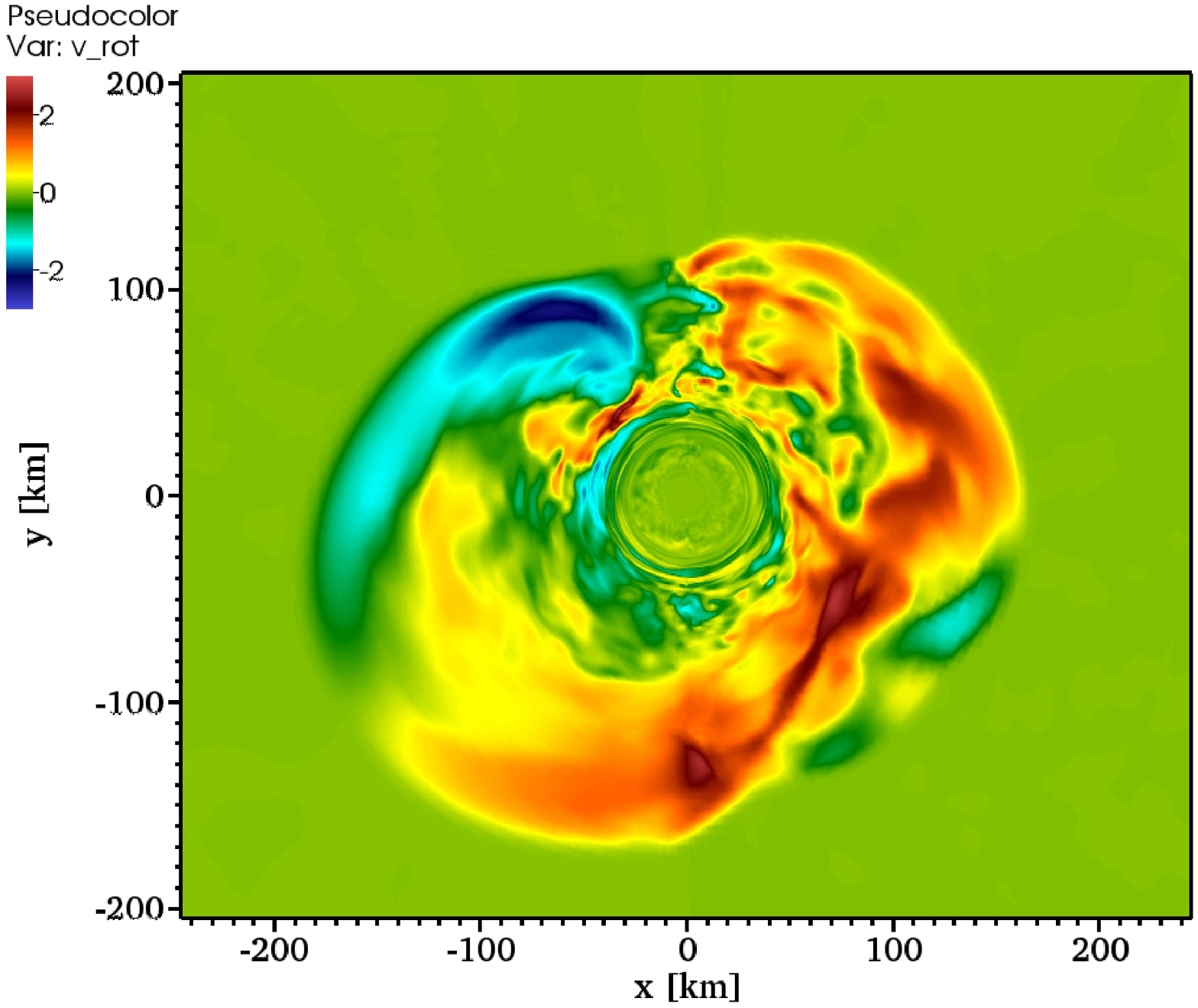}
  \caption{Snapshots of the rotational velocity around the origin in the
plane perpendicular to $\mathbf{n}=(-0.35, 0.93, 0.12)$ at post-bounce times
of 223\,ms, 227\,ms, 231\,ms, and 235\,ms during the 3D simulation of the
27\,$M_\odot$ progenitor.
Red and yellow (positive velocity values) correspond to counterclockwise
rotation.
    \label{fig:spiral}}
\end{center}
\end{figure*}

Quite remarkably, the SASI is not only able to reach larger amplitudes
in 3D than in 2D as long as its growth conditions remain favorable,
but it is even found to develop \emph{despite} some earlier convective
activity. Figure~\ref{fig:chi27} displays the critical parameter $\chi$
for the growth of convection as evaluated from spherically averaged
stellar quantities in the gain layer of our
3D simulation of the 27\,$M_\odot$ progenitor according to 
equation~(3) in \citet{mueller_12b},
\begin{equation}
\chi = \int_{\left\langle r_\mathrm{g}\right\rangle}^{\left\langle r_\mathrm{sh}\right\rangle}
\frac{\mathrm{Im}\,\omega_\mathrm{BV}}{|\left\langle v_r
\right\rangle |}\,\mathrm{d}r \,,
\label{eq:chi}
\end{equation}
where $\omega_\mathrm{BV}$ is the Brunt-V{\"a}is{\"a}l{\"a} frequency.
The integration is performed between the average gain
radius $\left\langle r_\mathrm{g}\right\rangle$ and the average shock 
radius $\left\langle r_\mathrm{sh}\right\rangle$. Note that only regions
contribute to the integral where $\omega_\mathrm{BV}^2 < 0$ indicates
local instability. The parameter $\chi$ roughly measures the ratio between
the advection timescale of the flow through the gain layer and the growth
timescale of convection. Since perturbations are advected out of the gain
layer with the accretion flow in a finite time, convection can develop only
when perturbations are amplified sufficiently strongly 
within this time. For the 
linear regime (i.e., for small initial perturbations) 
\citet{foglizzo_06} found the threshold condition of $\chi \gtrsim 3$ for convective
activity to develop in the accretion flow of the gain layer. This result
of mathematical analysis is supported by numerical studies in 2D by 
\citet{buras_06b,scheck_08,fernandez_09a,fernandez_09b}. 

Despite $\chi < 3$ (Figure~\ref{fig:chi27}), however,
convection develops around 80\,ms after bounce in our 3D simulation of 
the 27\,$M_\odot$ model. This happens because convective activity is
not only seeded by the artificially imposed, random density perturbations 
of 0.1\% amplitude (cf.\ Sect.~\ref{sec:setup1}) but also by
numerical perturbations along the axis of the computational
polar grid in one hemisphere, which we are not able to damp perfectly.
Although still small, these numerical effects are sufficiently large to 
trigger the rise of a buoyant plume against the advection flow, which
instigates further perturbations that exceed the linear regime.
In this case convection can be initiated although $\chi < 3$ signals
stability according to linear analysis \citep{foglizzo_06}. We note in passing that the level of seed perturbations was 
smaller and well compatible with linear theory in the 2D models
of \citet{mueller_12b}, and we also emphasize that the axial
perturbations have a noticeable effect only in the early growth phase of
convection whereas no significant axial artifacts or alarming flow 
perturbations near the polar axis can be observed
during the later phases of fully developed nonradial flow activity
in the postshock flow (see Figure~\ref{fig:3dsasi}). Once strong
SASI and convective mass motions have developed in the flow, the
supernova core contains a noise level so high that a subsequent decline 
of $\chi$ below the critical threshold of $\sim$3 does not imply that 
convective activity is unable to continue \citep{mueller_12b}.

Figure~\ref{fig:convection} shows a snapshot of the entropy in a 2D
slice at $152 \ \mathrm{ms}$. Here, the post-shock flow is still
dominated by multiple, intermediate-scale plumes as familiar from
buoyancy-driven convection. SASI shock sloshing becomes strong and
temporarily dominant only afterward. The argument that any
convective activity arising from sufficiently large seed perturbations
will quench the SASI thus seems to be invalid. The competition between
convection and the SASI is obviously more subtle than a superficial
reading of recent papers
\citep{burrows_12,murphy_12,mueller_12b,ott_12} might suggest. Some
relevant aspects of the competing growth conditions and interaction 
of the two instabilities were discussed on the basis of 2D supernova
simulations by \citet{scheck_08}. A very similar behavior can be
diagnosed in the 3D case and will be addressed on the basis of 
parametric 3D studies in Sect.~\ref{sec:para25sunmodel}, where we
will make use of an axis-free Yin-Yang grid in order to avoid the 
perturbative influence of the polar coordinate axis during the linear
growth phase of the seed perturbations.

\begin{figure}
\begin{center}
% \plotone{angular_momentum_gain_region.eps}
  \includegraphics[width=\columnwidth]{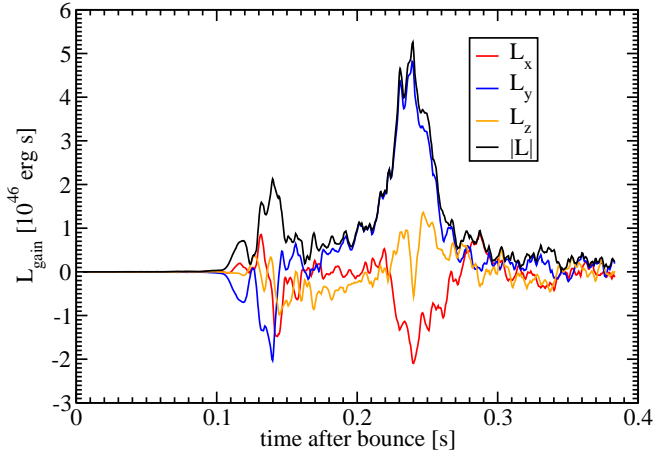}
  \caption{Time evolution of the angular momentum components
    $L_x$, $L_y$, and $L_z$, and of the absolute magnitude
    $|L|$ of the total angular momentum contained in the gain region
    of the 27\,$M_\odot$ simulation in 3D.
    \label{fig:angular_momentum}}
\end{center}
\end{figure}

\subsection{Detecting the spiral mode of the SASI}
While the SASI is limited to a sloshing motion along the symmetry axis
in 2D, there is the possibility of a spiral mode in 3D
\citep{blondin_07,iwakami_09,fernandez_10}, which could provide a means for
angular momentum separation between the PNS and the
ejecta, and might also have different saturation properties in the
non-linear phase.

Detecting the spiral mode is not straightforward, as the time
evolution of the coefficients $a_\ell^m$ needs to be taken into
account. Merely computing the coefficients with $m \neq 0$ is
\emph{not} sufficient. In principle, spiral and sloshing modes can be
disentangled by a Fourier analysis of $a_\ell^m (t)$
\citep{iwakami_08} if they remain stable over several oscillation
periods.  We use a somewhat different approach to visualize
the character of the $\ell=1$ mode in our 3D model here. The
coefficients $a_1^m$ can be combined into a vector,
\begin{equation}
\mathbf{a}_1 = 
(a_{1}^{1}, a_{1}^{-1}, a_{1}^{0}) =
(a_x, a_y, a_z),
\end{equation}
which is a rough measure of the angle-averaged displacement 
of the shock center from the
origin and also indicates the direction and amplitude of the shock
deformation (neglecting modes with higher $\ell$). We visualize the
time evolution of this amplitude vector in 3D space in
Figure~\ref{fig:3d_amplitude} with different colors of the trajectory
indicating different phases of the simulation.

Once the SASI starts to grow vigorously (blue curve), $\mathbf{a}_1$
initially moves along a narrow elliptical path with growing semi-major
axis, indicating a predominant sloshing mode. Towards the phase of
strongest SASI activity, the trajectory becomes more circular,
signaling the transition to a spiral mode. The plane of the spiral
remains relatively stable until the maximum amplitude is
reached and the SASI dies down again. It is roughly perpendicular to
the vector $\mathbf{n}=(-0.35, 0.93, 0.12)$, i.e.\ there is no
alignment with the axis of the spherical polar
grid. Figure~\ref{fig:3d_amplitude} also further illustrates the
different behavior during the ``SASI-dominated'' phase compared to the
earlier and later ``convectively-dominated'' phases, during which
$\mathbf{a}_1$ evolves in a more random fashion.

Slicing the model along the plane in which $\mathbf{a}_1$
predominantly moves allows us to visualize the distinctive spiral mode
pattern, as shown in Figure~\ref{fig:spiral}. The snapshots of the
rotational velocity around the ``axis'' of the spiral mode
(in the plane through the origin perpendicular to that axis)
reveal two
counter-rotating regions. While these regions are initially of
comparable size, the flow in the counter-clockwise direction
eventually dominates, and the rotation of the mode pattern with a
continuously shifting triple point can clearly be seen.

During the short phase of strong SASI activity, angular momentum
separation by the spiral mode proceeds very efficiently, transferring
a total angular momentum of $\sim 5 \times 10^{46} \ \mathrm{erg}
\ \mathrm{s}$ into the gain region
(Figure~\ref{fig:angular_momentum}). As expected, the direction of the
angular momentum vector is extremely similar to the normal vector of
the spiral plane. 
However, all the angular momentum is eventually advected out of the
gain region after the SASI has died down.

%--------------------------------------------------------------------------

\section{Parametric 2D and 3D post-bounce simulations of a
25\,$M_\odot$ star}
\label{sec:para25sunmodel}

In order to probe the SASI growth conditions and the differences
between 2D and 3D accretion flows in more detail, we performed a set
of simulations for the 25\,$M_\odot$ solar-metallicity progenitor of 
\citet{woosley_02}. The choice of this stellar model is motivated
by the fact that it turned out to be particularly hard to explode 
in self-consistent 2D simulations with sophisticated neutrino treatment,
where it exhibits violent shock sloshing motions with amazingly stable
periodicity over a long period of postbounce evolution 
\citep{mueller_13,janka_12b}.
It thus offers promising perspectives to explore the relevance of 
the SASI in 3D for a second progenitor in addition to the 27\,$M_\odot$
case investigated in Sect.~\ref{sec:27msunmodel}. 
Because of computer time constraints we have 
performed the parametric study with simplified, gray neutrino transport
only for the 25\,$M_\odot$ progenitor, but we did not repeat such a 
study for the 27\,$M_\odot$ star. Since the crucial parameters
that regulate the postbounce dynamics of the stalled shock and accretion
layer (in particular the neutron-star contraction rate and the 
core-neutrino luminosities) can be set as boundary conditions, it must
be expected that the basic features observed in the 25\,$M_\odot$
simulations, especially the different regimes of postshock 
instabilities, can also be found for the 27\,$M_\odot$ model for 
appropriate choices of the parameter values.

\subsection{Numerical methods and modeling setup}
\label{sec:setup2}

In order to efficiently run several 3D simulations, a
simplified setup is employed that was extensively applied before by 
\citet{wongwathanarat_10,wongwathanarat_12,mueller_e_12,
scheck_06,scheck_08,arcones_07,arcones_11}. 
Instead of the elaborate neutrino transport solver used
for the 27\,$M_\odot$ run in Sect.~\ref{sec:27msunmodel}, now 
an approximate, gray (ray-by-ray) neutrino transport is used
(for details, see \citealp{scheck_06}). Moreover,
the high-density interior of the proto-neutron star (at densities 
above neutrino optical depths between some 10 and several 100, depending
on the phase of the evolution) is excised and replaced by an inner 
grid boundary (also in contrast to the 27\,$M_\odot$ run, where the
whole neutron star was included in the simulation). 
A contraction of the boundary radius is prescribed to mimic
the shrinking neutron-star core \citep{scheck_06,arcones_07},
and neutrino luminosities of chosen magnitude
and time dependence are imposed there to parametrize
the neutrino losses from the interior core volume. The transport 
approximation allows us to account for radial flux variations due to
accretion luminosity and neutrino energy deposition in the gain layer.
Although this modeling approach is not elaborate enough to exactly 
reproduce the results obtained with more sophisticated and fully
consistent treatments, it nevertheless includes a reasonably good
representation of all aspects that determine the flow dynamics in 
collapsing stellar cores. The parametrized approach has the big
advantage that different components can be controlled 
to a large extent independently
so that studies with varied conditions can be performed in a 
systematic manner.

The computational efficiency is further enhanced by the use of an
axis-free Yin-Yang grid \citep{wongwathanarat_10b},
which avoids the time-step constraints and numerical artifacts
close to the axis of the polar grid. The 3D runs were conducted
with a grid zoning of $600(r)\times47(\theta)\times137(\varphi)\times2$,
the corresponding 2D models with $600(r)\times90(\theta)$ zones, which
ensures an angular resolution of $2^\circ$ in both 2D and 3D. The
inner grid boundary is placed at an enclosed mass of approximately 
1.1\,$M_\odot$. When the models are started from 1D conditions at 
15\,ms after bounce, the corresponding initial radius of the inner
grid boundary, $R_{\mathrm{ib}}^{\mathrm{i}}$, is located at 53\,km,
while the radius of the outer grid boundary, $R_{\mathrm{ob}}$, is
chosen to be at 10000\,km in the beginning. The radial grid resolution, 
$\Delta r$, is kept constant at 0.265\,km up to the radius of 
initially 100\,km, beyond which the zone width is increased 
logarithmically.

The shrinking core of the nascent neutron star is described by the
contraction of the inner grid boundary according to equation~(1) of
\citet{scheck_06} with an exponential timescale of
$t_{\mathrm{ib}} = 1$\,s and a final radius of 
$R_{\mathrm{ib}}^\mathrm{f}$. Along with the retraction of the inner
grid boundary the whole radial grid is moved with the same velocity.
During the computed model evolution the neutrino luminosities 
imposed at the inner grid boundary are kept constant. The mean
neutrino energies of the neutrino fluxes entering the grid are
prescribed as functions of the temperature in the first radial cell
of the computational grid as described by \citet{ugliano_12}.

In order to trigger the growth of nonradial hydrodynamic instabilities,
seed perturbations with randomly varied amplitude are imposed as 
zone-to-zone variations of the radial velocity in the whole 
computational domain at the start of the 2D or 3D simulations.

We present results of three types of models, which differ in the 
size of the seed perturbations, either of 0.1\% or 3\% 
amplitude (``small'' and ``large'' initial perturbations, 
respectively), and in the contraction behavior adopted for the 
neutron-star core. The latter is controlled by the assumed
asymptotic radius of the inner grid boundary at $t\rightarrow\infty$,
namely $R_{\mathrm{ib}}^\mathrm{f} = 15$\,km for ``slow contraction''
and $R_{\mathrm{ib}}^\mathrm{f} = 10$\,km for ``fast contraction''.
In the case of slow contraction the imposed boundary luminosities
of $\nu_e$ and $\bar\nu_e$ are $1.49\times 10^{51}$\,erg\,s$^{-1}$ 
and $0.89\times 10^{51}$\,erg\,s$^{-1}$, respectively; in the case
of fast contraction we use 
$L_{\nu_e,\mathrm{ib}} = 5.96\times 10^{51}$\,erg\,s$^{-1}$ and
$L_{\bar\nu_e,\mathrm{ib}} = 3.58\times 10^{51}$\,erg\,s$^{-1}$.
(Muon and tau neutrinos and antineutrinos are included as well but
do not play an important role for the hydrodynamics of the accretion
flow in the supernova core.)
The three computed models are characterized by the following 
parameter sets:
\begin{enumerate}
\setlength{\itemindent}{1em}
\item[\bf S25-1:] slow boundary contraction, 3\% perturbation amplitude;
\item[\bf S25-2:] fast boundary contraction, 0.1\% perturbation amplitude;
\item[\bf S25-3:] fast boundary contraction, 3\% perturbation amplitude.
\end{enumerate}

Model S25-1 was evolved until 515\,ms after bounce, while S25-2 and
S25-3 were run until 365\,ms after core bounce.

\begin{figure*}
\begin{center}
% \plotone{2d3dcomp.eps}
  \includegraphics[width=\textwidth]{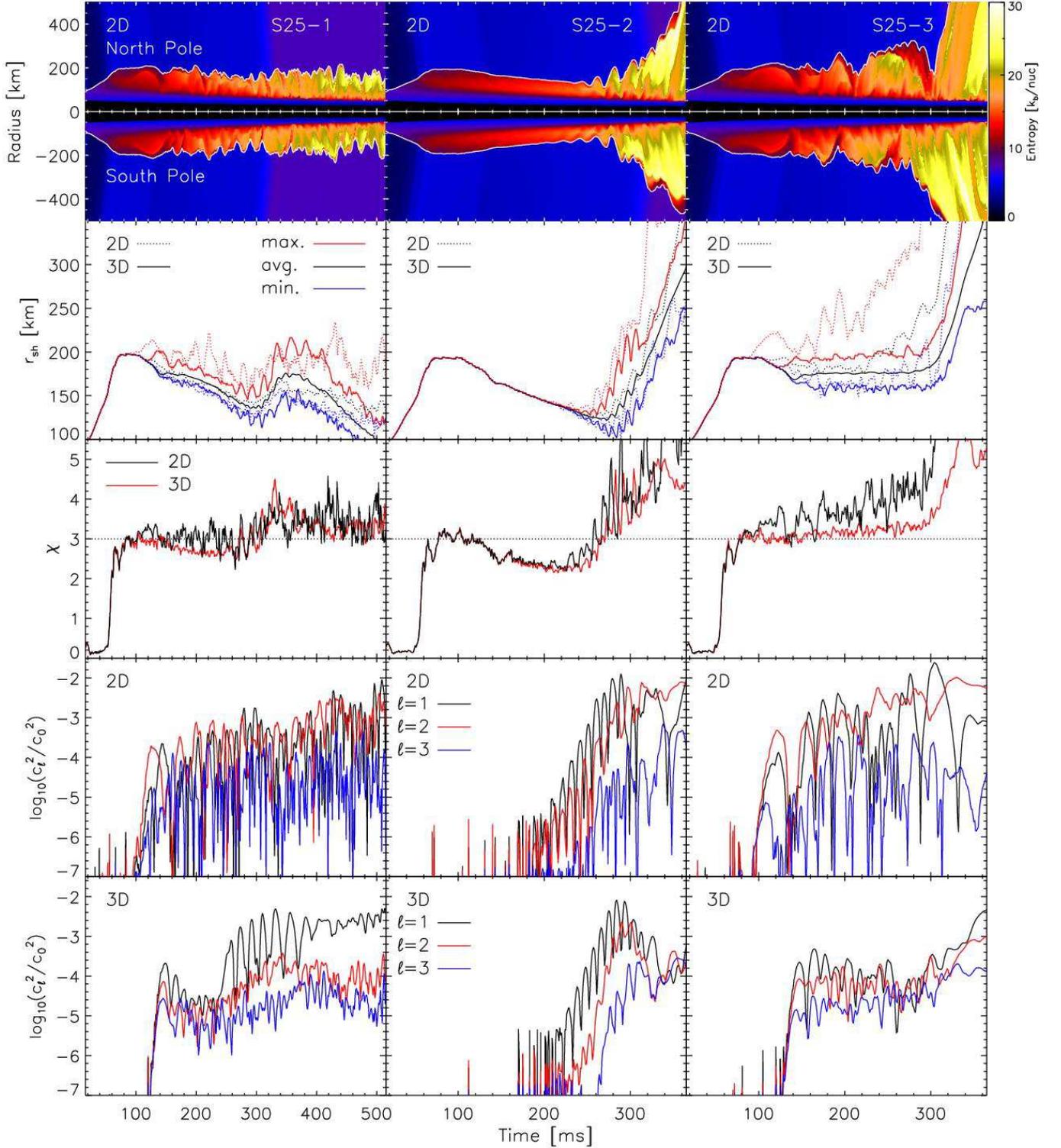}
  \caption{Comparison of the time evolution of 2D and 3D results of our
    set of parametrized postbounce simulations S25-1, S25-2,
    and S25-3 of the 25\,$M_\odot$ progenitor.
    The upper panels show north polar and south polar
    entropy profiles for the 2D runs with thin white lines indicating
    the shock trajectories, which separate low-entropy preshock matter
    (black and blue) from the high-entropy postshock region (red, orange,
    yellow). The second panels from the the top
    display maximum, average, and minimum shock radii for 2D (dotted)
    and 3D models, the panels in the third line present the $\chi$
    parameter for convective instability (Equation~\ref{eq:chi}) of 2D (black
    lines) and 3D (red lines) cases, and
    the plots in the fourth and fifth lines provide the
    normalized pseudo-power coefficients $c_\ell^2/c_0^2$
    (Equation~\ref{eq:cl}) on a logarithmic scale for the 2D and 3D
    simulations, respectively. Phases with values of $\chi \lesssim 3$
    correlate with preferred growth of low-mode ($\ell = 1,\,2$) SASI
    activity.
    \label{fig:2d3dcomp}}
\end{center}
\end{figure*}

\begin{figure*}
\begin{center}
\includegraphics[width=.40\textwidth]{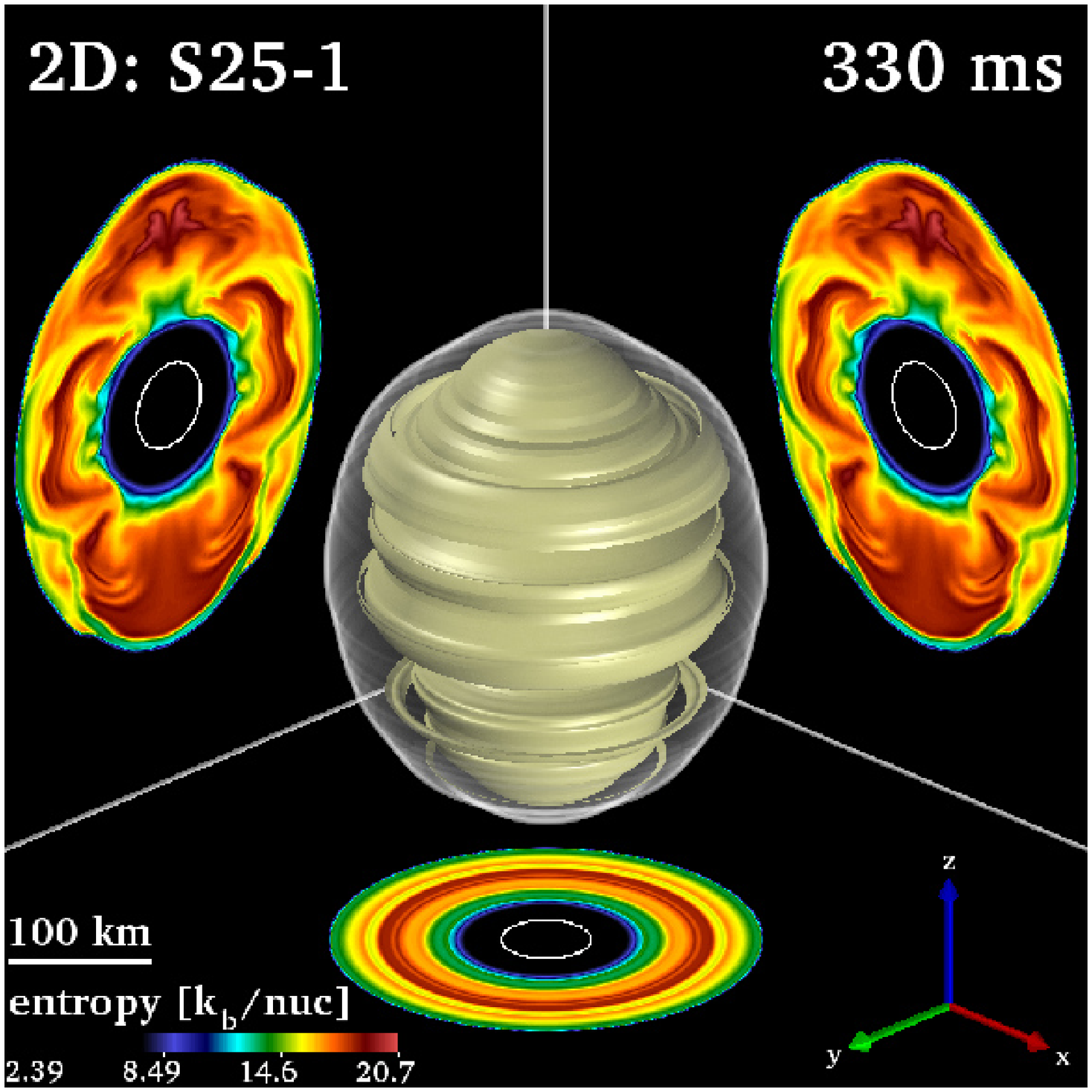}
\includegraphics[width=.40\textwidth]{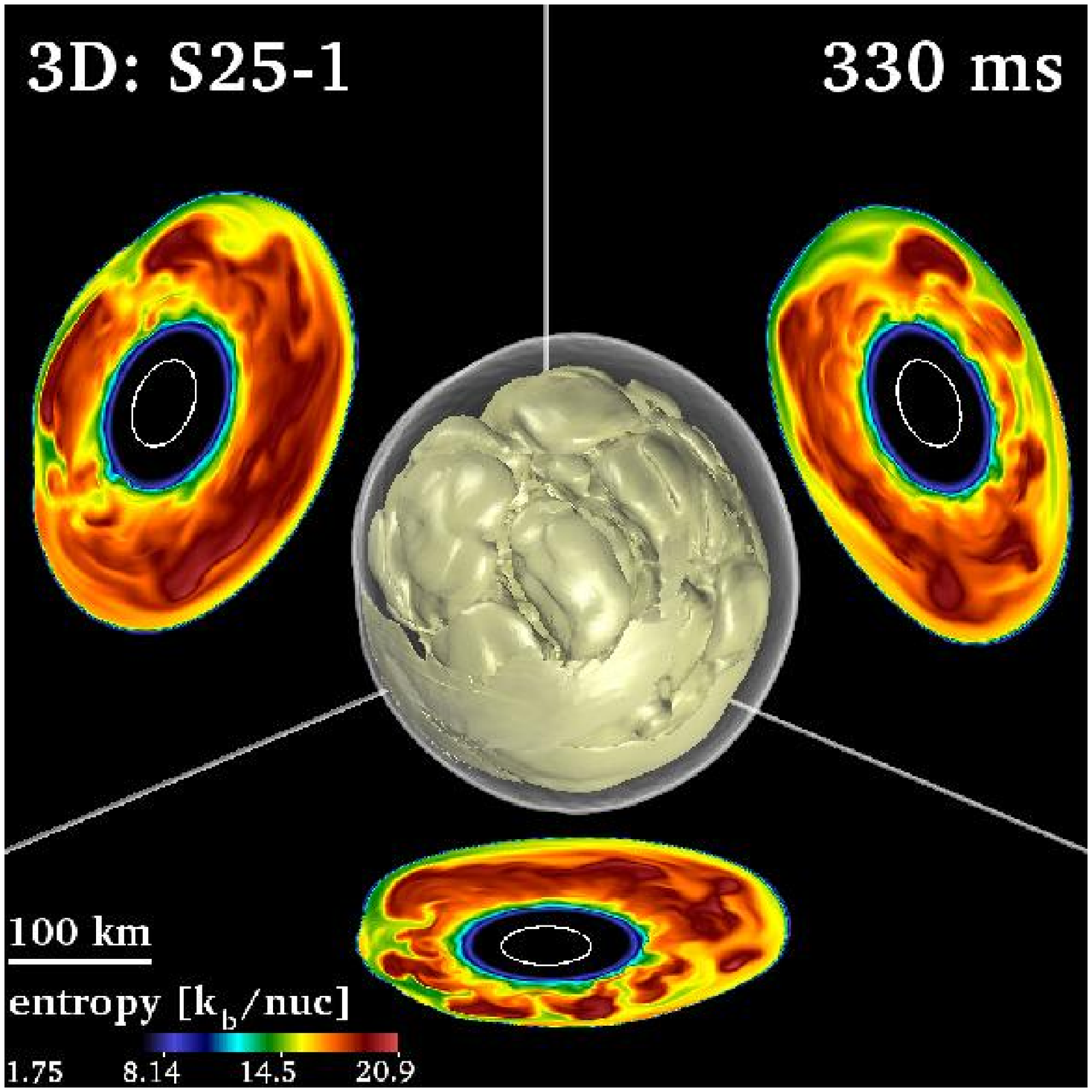}\\
\includegraphics[width=.40\textwidth]{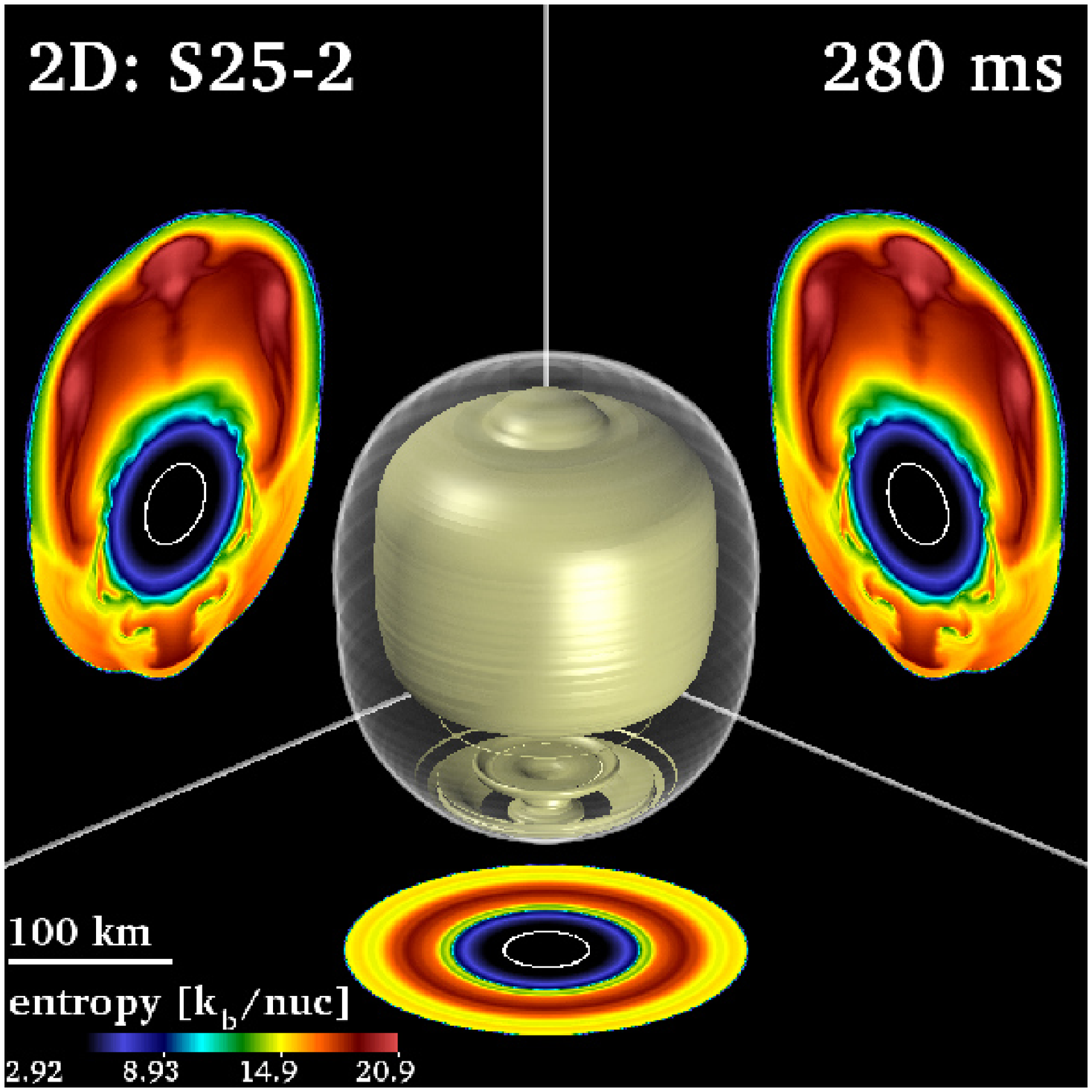}
\includegraphics[width=.40\textwidth]{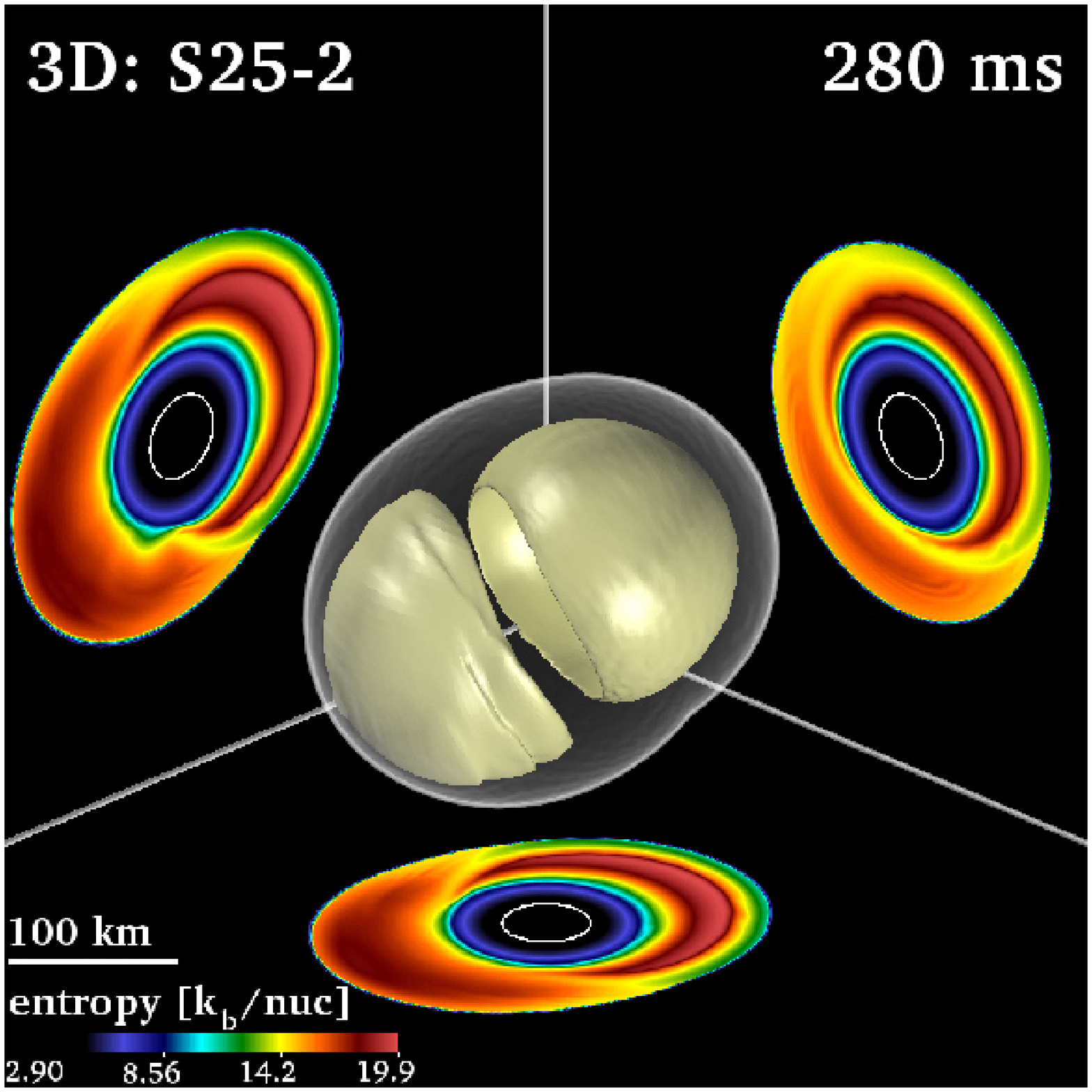}\\
\includegraphics[width=.40\textwidth]{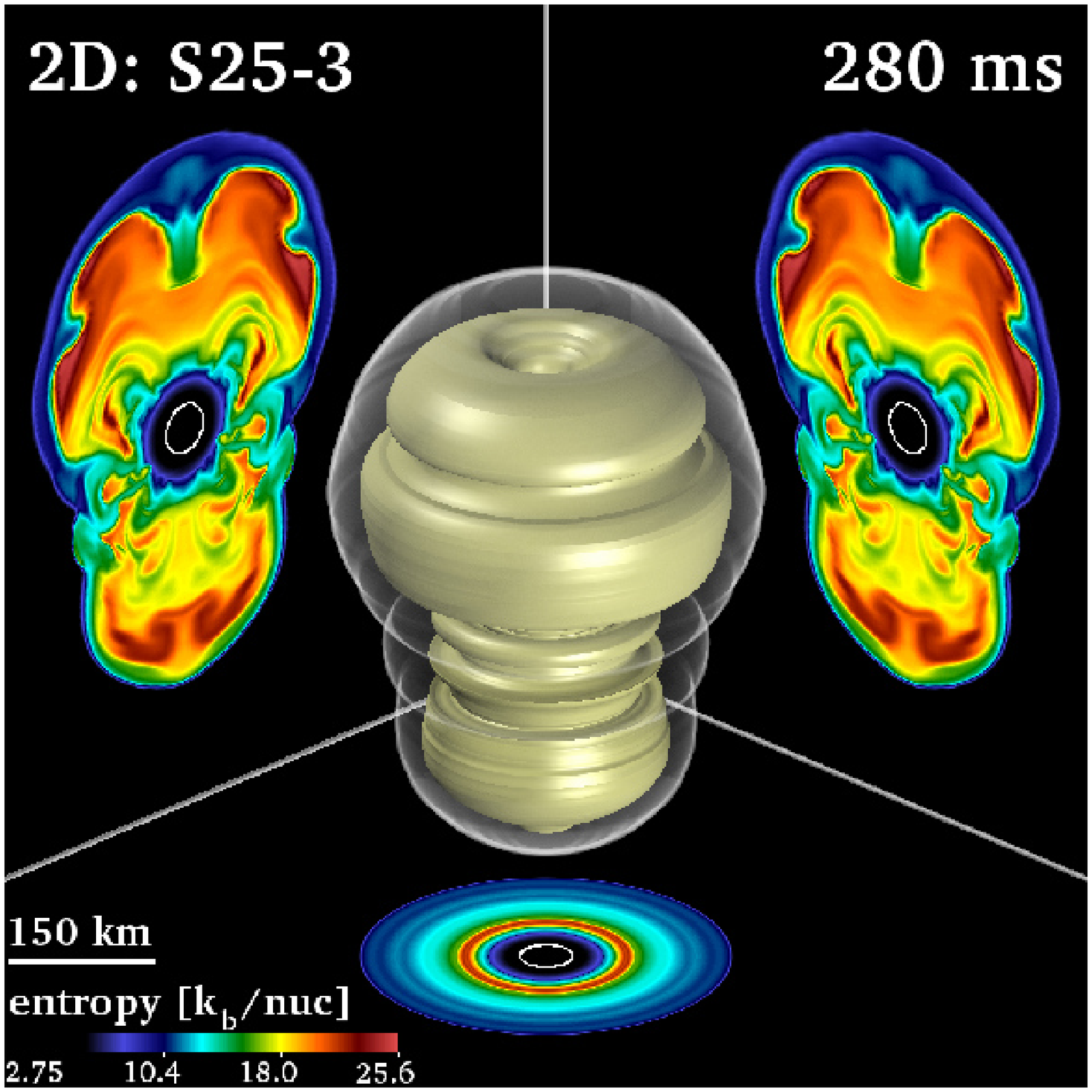}
\includegraphics[width=.40\textwidth]{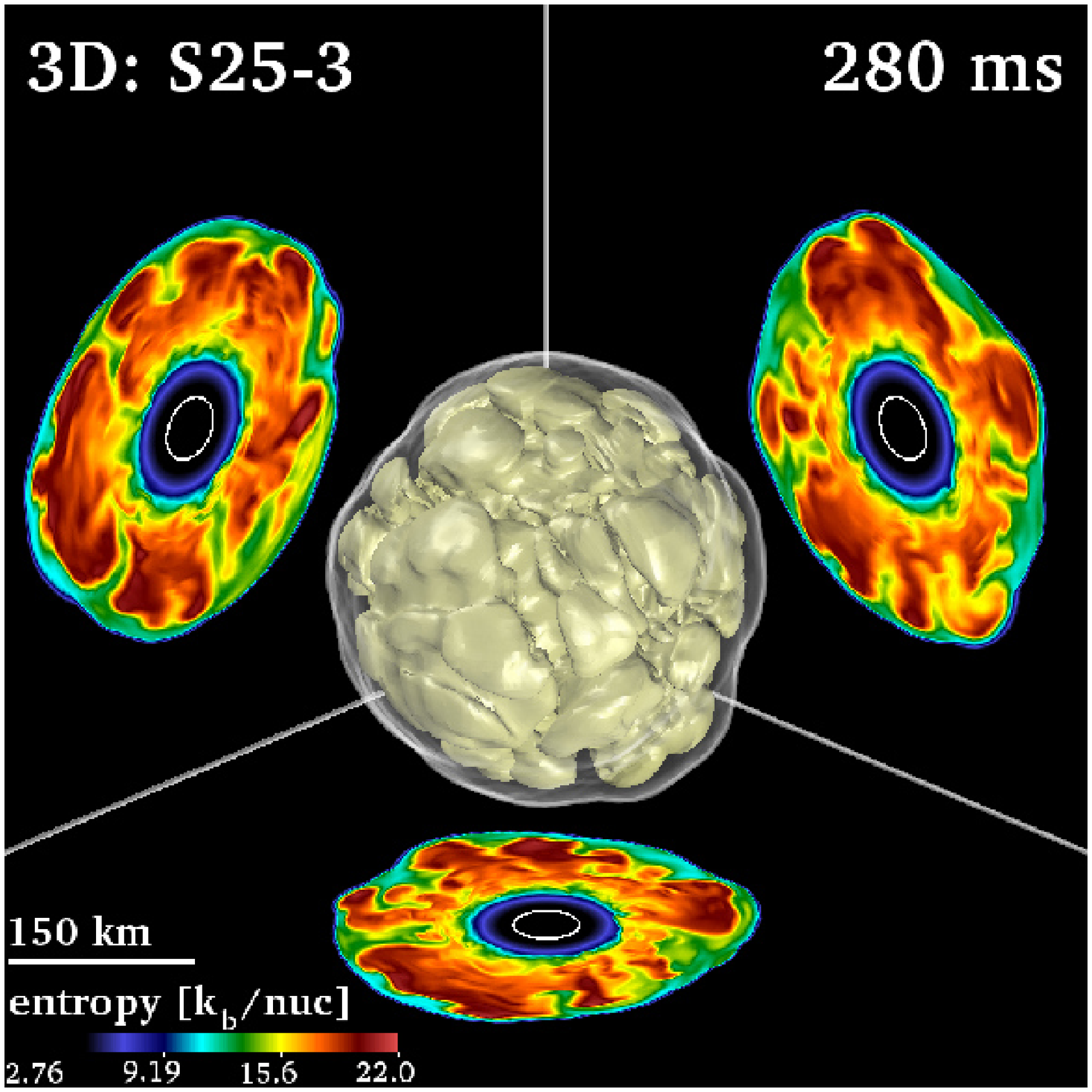}
\caption{\label{fig:2d3dsasi}
   Structure of the 2D ({\em left}) and 3D ({\em right}) models
   of our set of parametrized 25\,$M_\odot$ simulations
   ({\em top:} S25-1; {\em middle:} S25-2; {\em bottom:} S25-3)
   at representative postbounce times (as given in the upper
   right corner of each panel). The central object in all
   panels provides a three-dimensional visualization of the
   supernova shock (white, transparent surface), which engulfs
   a surface of constant entropy of 17\,$k_\mathrm{b}$ per
   nucleon. The images on the cube walls show entropy
   distributions (the color coding as given by the color bars
   in the lower left corners with black, blue, and green
   signaling low values) in the $x$-$y$, $x$-$z$, and
   $y$-$z$ planes (orientation according to triad in the
   lower right corner of the upper left panel). The white
   circle indicates the inner grid boundary and a scale stick
   in each panel gives a measure of the size of the displayed
   volume. While the 3D model of S25-1 exhibits signatures of
   neutrino-driven convection and of low-order
   multipole asymmetry due to SASI shock sloshing, both the
   2D and 3D cases of S25-2 show a clear dominance of SASI
   activity, whereas model S25-3 possesses the clearest pattern
   of neutrino-driven buoyancy.
  }
\end{center}
\end{figure*}

\subsection{SASI activity in dependence of supernova-core
conditions}
\label{sec:sasivariation}

Figure~\ref{fig:2d3dcomp} 
provides an overview of the dynamical evolution
of the three models S25-1, S25-2, and S25-3 in 2D compared to 3D
by showing the north polar and south polar entropy profiles
of the 2D simulations in the top panels, shock radii (average,
maximum, minimum) in the second line from the top, the critical
$\chi$ parameter (evaluated according to Equation~\ref{eq:chi}) in
the panels of the third line, and a spherical harmonics mode
decomposition of the deformed shock surface in the fourth 
and fifth lines for 2D and 3D cases, respectively.

Here we follow the deformation mode analysis used
by \citet{wongwathanarat_12}
and write the time-dependent decomposition of the 3D shock surface,
$r_{\mathrm{sh}}(\theta,\varphi)$, in spherical harmonics as
\begin{equation}
\label{eq:ylmexpansion}
r_{\mathrm{sh}}(\theta,\varphi)=\sum_{\ell=0}^\infty\sum_{m=-\ell}^\ell
b_\ell^mY_\ell^m(\theta,\varphi)\, ,
\end{equation}
where $b_\ell^m$ are the expansion coefficients and the spherical
harmonics $Y_\ell^m$ are functions of
the associated Legendre polynomials $P_\ell^m$,
\begin{equation}
\label{eq:ylm}
Y_\ell^m(\theta,\varphi)=K_\ell^mP_\ell^m(\cos{\theta})e^{im\varphi}\, ,
\end{equation}
with
\begin{equation}
\label{eq:klm}
K_\ell^m=\sqrt{\frac{2\ell+1}{4\pi}\frac{(\ell-m)!}{(\ell+m)!}}\, .
\end{equation}
Multiplying Equation~(\ref{eq:ylmexpansion}) by the complex conjugate of the
spherical harmonic, $Y_\ell^{m*}$, and integrating over the solid angle,
the expansion coefficient is found to be
\begin{equation}
\label{eq:clm}
b_\ell^m=\int_0^{2\pi}d\varphi\int_0^\pi
d\theta\;\sin{\theta}\,r_{\mathrm{sh}}(\theta,\varphi)Y_\ell^{m*}(\theta,\varphi)\,.
\end{equation}
For our mode analysis we actually use the
pseudo-power coefficients $c_0^2 \equiv |b_0^0|^2$ and
\begin{equation}
 c_\ell^2 \equiv \frac{1}{2\ell+1} \left( |b_\ell^0|^2 +
                                  2 \sum_{m=1}^{m=\ell} |b_\ell^m|^2 \right)
\label{eq:cl}
\end{equation}
for $\ell>0$, respectively, whose values do not depend on
the particular choice of the orientation of the coordinate grid in 3D.
The $c_\ell^2$ are related to the coefficients $a_\ell^m$
of the real spherical harmonics decomposition in
Equation~(\ref{eq:ylm_decomposition}) by
\begin{equation}
 c_\ell^2 \equiv 4 \pi \sum_{m=-\ell}^{m=\ell} |a_\ell^m|^2.
\end{equation}
In Figure~\ref{fig:2d3dcomp} logarithmic values of $c_\ell^2$ normalized by
$c_0^2$ are plotted.

The north polar and south polar entropy profiles of the 2D simulations
in the upper panels of Figure~\ref{fig:2d3dcomp} show that the slow
contraction case, S25-1, does not explode within 500\,ms after bounce,
whereas the fast contraction models
S25-2 and S25-3 develop explosions around 300\,ms post bounce 
because of the stronger neutrino heating by higher boundary and,
in particular, accretion luminosities. The dynamics
of the shock and postshock flow until the explosions set in reveals 
interesting differences in all three cases.

The north and south polar shock trajectories as well as the maximum
and minimum shock radii (given by dotted lines in the panels on 
the second line of Figure~\ref{fig:2d3dcomp}) in all three 2D cases
exhibit pronounced sloshing motions, which suggest considerable 
activity on the scale of low-order spherical harmonics modes.
The most regular pattern of alternating north-polar and south-polar
shock expansions and contractions, however, is visible in model S25-2,
whose periodic shock excursions remind one strongly of the clean SASI 
sloshing mode found for the 27\,$M_\odot$ progenitor by 
\citet{mueller_12b}. Before large-amplitude shock motions
are reached, the sloshing amplitude grows slowly in an otherwise
seemingly quiet environment. A similar behavior was observed by
\citet{scheck_08} in their 2D models W00 and W00F, whose 
parameter settings (chosen core contraction rate in combination
with very low core-boundary luminosities) were initially disfavorable
for the onset of neutrino-driven convection and thus allowed one to
follow the SASI growth from the linear to the nonlinear regime in a 
very clear way.

In contrast, the 2D models of S25-1 and S25-3 show shorter-timescale 
variability of the shock trajectories superimposed on a 
longer-timescale quasiperiodic north-south sloshing of the shock.
This suggests the presence of asymmetries of higher-order spherical 
harmonics type as they are typical of neutrino-driven buoyant plumes.

These descriptions are supported by a more quantitative analysis
of the growth conditions of convective instability in the 
postshock accretion layer as expressed by the critical $\chi$
parameter, which measures the ratio of average advection timescale
of the accretion flow through the gain region to the convective growth
timescale. Inspecting
the panels in the third line of Figure~\ref{fig:2d3dcomp}, where the
black curves display the 2D results, reveals that only in model 
S25-2 $\chi$ drops below the critical threshold value of 3 for 
most of the postbounce evolution until strong shock expansion and
explosion sets in. The low values of $\chi$ are a consequence of 
the strong shock contraction in response to the rapidly shrinking 
core of the neutron star assumed in this model. As expected from the 
linear perturbation analysis of \citet{foglizzo_06}, 
which is applicable to model S25-2 because of the small amplitude
of the initial seed perturbations, convective overturn cannot develop
and the model exhibits the observed clean SASI growth pattern.
Correspondingly, the spherical harmonics decomposition of the 
shock deformation shows the characteristic oscillatory growth of 
dominant $\ell = 1$ (dipole) and $\ell = 2$ (quadrupole) modes
to nonlinear magnitudes over a timescale of roughly 100\,ms (fourth
panel from top in the middle column of Figure~\ref{fig:2d3dcomp}). 

In contrast, the 2D models of S25-1 and S25-3 show a growth of the
shock asymmetries that differs from that of S25-2.
A long-time oscillatory rise of the 
amplitude is not visible in both cases, but relatively large shock 
deformation develops much earlier and the time dependence of the
spherical harmonics amplitudes of order $\ell = 1,\,2,\,3$ is
much more irregular than in S25-2. Although still dipolar and 
quadrupolar shock deformation modes dominate, 
the short-timescale variability 
suggests bubbles and plumes on the smaller angular scales
typically connected to the occurrence of strong buoyancy activity in 
the neutrino-heating layer. This is in agreement not only with $\chi$
values around the critical number of 3 for S25-1 and values well
above 3 for S25-3, but it is also compatible with the visual
impression of the entropy structures between shock and gain radius
displayed in the left panels of Figure~\ref{fig:2d3dsasi}.
The pronounced deformation along the symmetry axis, which is 
responsible for the dominance of the dipolar mode in many evolution
phases of all three models, can be nicely seen in the shape of the 
shock and of the surface of constant entropy displayed by the 
central, three-dimensional representation of the 2D model structure
in each of the left panels. It should be noted, however, that different
from the situation of S25-1 and S25-3, where the entropy surface 
shows short-wavelength variations reflecting the presence of buoyant
plumes, this surface is much smoother in model S25-2, where the
violent SASI sloshing at 280\,ms post bounce is accompanied by 
much weaker convection. 

The rapid onset of buoyancy in models S25-1 and S25-3 is fostered
by the larger initial seed perturbations. Since the difference
between S25-2 and S25-3 is {\em only} the size of the amplitude of
the initial random seeds (0.1\% for S25-2 and 3\% for S25-3),
the different evolution of the flow asymmetries and the different
values of $\chi$ in these two models clearly demonstrate the 
important influence of the magnitude of preexisting deviations
from spherical symmetry in the collapsing stellar core.

A comparison of 2D and 3D results for the same modeling setup
is extremely interesting now. Inspecting the right panels
of Figure~\ref{fig:2d3dsasi} it becomes obvious that the SASI
activity appears in a much clearer way in the entropy
surfaces of the 3D models. The shock sloshing creates smooth, 
coherent hemispheric shells of constant entropy like those we
have seen in Figure~\ref{fig:3dsasi} and already discussed in 
Sect.~\ref{sec:results}. The 3D case of S25-2 exhibits such
bipolar hemispheric entropy lobes, which are nearly axisymmetric 
around a direction tilted relative to the three coordinate
axes (indicated by the triad), but which possess hardly any 
substructure on shorter wavelength scales. This indicates that the 
entropy variations are created by pure SASI sloshing movements
of the shock. The three $x$-$y$, $x$-$z$, and $y$-$z$ cuts 
displayed on the cube walls yield an impressive confirmation of
nearly perfect entropy half-shells without any indication of
secondary, parasitic instabilities of Rayleigh-Taylor or
Kelvin-Helmholtz type. In contrast, model S25-3 shows the
familiar cauliflower pattern of expanding and fragmenting 
Rayleigh-Taylor fingers and Kelvin-Helmholtz swirls typical
of neutrino-driven buoyancy. Model S25-1 is again different and
characterized by features of both types in superposition: The
snapshot of the upper right panel in Figure~\ref{fig:2d3dsasi}
displays a convective bubble pattern that is engulfed by a 
smooth half-shell closer behind the shock in one hemisphere.

These differences are reflected by the spherical harmonics
amplitudes for the $\ell = 1,\,2,\,3$ modes of the shock 
deformation in the 3D models presented in the bottom panels
of Figure~\ref{fig:2d3dcomp}. While in models S25-1 and S25-2
the $\ell = 1$ deformation around the times of the snapshots 
in Figure~\ref{fig:2d3dsasi} (330\,ms and 280\,ms, respectively) 
remains clearly dominant for longer evolution
phases, and the time variation of the dipole amplitude
reveals very stable periodicity for 100\,ms and more, all
three lowest-order spherical harmonics have similar strength
in model S25-3 around 280\,ms after bounce. In S25-3 the dipolar 
asymmetry component has the largest amplitude only for 
some short episodes of the postbounce evolution, although
quasi-periodic $\ell = 1$ amplitude variations can 
be observed also in this case. The higher-order multipoles
in all models show greater variation frequencies and less good
temporal regularity and seem to be 
influenced by the stochasticity associated with convective 
mass motions.

The $\chi$ parameter (panels in the third line of 
Figure~\ref{fig:2d3dcomp}) aids a better understanding of the 
differences between the 3D models and of their differences
relative to the corresponding 2D results. In the 2D and 3D
cases of model S25-2 the values of the $\chi$ parameter evolve
essentially identically until $\sim$250\,ms and stay below
the critical threshold of 3 for convection for most of the 
time. Correspondingly,
both the 2D and 3D simulations do not develop any significant
level of convective activity. Therefore pure SASI activity can
be observed, which in both cases grows in a very similar way 
and is characterized by a dipolar shock deformation mode
that amplifies most rapidly and that dominates the 
$\ell = 2,\,3$ spherical harmonics components in 3D even more
than in 2D.

In model S25-3, which differs from S25-2 only in the size of
the initial seed perturbations, the quick appearance of 
buoyancy asymmetries leads to a stronger shock expansion,
which prevents $\chi$ from dropping below 3 in both the 2D 
and 3D runs. In this case SASI activity cannot be unambiguously
diagnosed by persistent shock sloshing or spiraling. The 
amplitudes of the $\ell = 1,\,2,\,3$ shock deformation modes
in 3D remain fairly small (much smaller than the maximum amplitudes
of S25-1 and S25-2), and a clear hierarchy of these lowest-order
multipoles over longer periods of time is not established in
the 3D case of S25-3 (Figure~\ref{fig:2d3dcomp}, bottom right panel).

Models S25-2 and S25-3 in comparison confirm that the growth 
conditions of SASI activity improve in phases of strong shock
contraction. As discussed in \citet{scheck_08}, the
SASI growth rate roughly scales inversely with the advection
timescale of the accretion flow from the shock to the neutron-star
surface. In model S25-3 the stronger shock expansion, triggered
by convection that is seeded by the (nonlinear) initial 
perturbations, leads to longer advection times and thus disfavors
the development of the SASI. 

Model S25-1 demonstrates another interesting effect. The SASI can
be stronger in 3D models than in the corresponding 2D cases. In the 
3D version of S25-1 the shock deformation mode develops a very 
large $\ell = 1$ amplitude, for which after 250\,ms $c_1^2$ exceeds
the amplitudes of the higher-order multipoles by more than an order
of magnitude (Figure~\ref{fig:2d3dcomp}, bottom left panel). While
quasiperiodic shock oscillations with nice regularity and nearly
stable axis characterize the evolution until about 350\,ms, the 
direction of the dipole axis wanders afterward to give way to a
strengthening spiral mode after $\sim$500\,ms, which coincides with 
the phase of most extreme shock recession. In the 3D case the
SASI activity appears much more strongly than in 2D, where dipolar 
and quadrupolar deformation modes have similar size and higher 
multipoles, associated with convective mass motions, contribute
significantly. The relevance of convection in the 2D model is
compatible with the $\chi$ being around 3 in this case, whereas
$\chi < 3$ in the 3D run reduces the influence of buoyancy but
signals more favorable conditions for the SASI because of a 
shorter advection timescale (similar to the situation in S25-2).

It is important to note that both S25-1 and S25-3 were computed 
with large initial seed perturbations. In the case of S25-3,
however, the larger boundary and accretion luminosities lead to
stronger neutrino heating, which supports buoyancy and allows the
shock radius to stay large even in the 3D case. On the contrary,
in S25-1 the weaker neutrino heating does not prevent shock
retraction.

Convection and SASI activity in 3D therefore depend sensitively 
on the behavior of the accretion shock, which in turn reacts to
the neutron star contraction and the power of neutrino-energy
deposition. Our set of models shows that for a favorable
combination of conditions the SASI in 3D can be much stronger
(S25-1) and purer (S25-2) than in the corresponding 2D cases.
In general, however, the development of convection and SASI
in the postshock accretion layer seems to be fully compatible
with the general theoretical understanding obtained in 
connection of 2D hydrodynamic simulations in \citet{scheck_08}.

\section{Discussion and Conclusions}
\label{sec:conclusions}
We have simulated the post-bounce evolution of the 27\,$M_\odot$
progenitor of \citet{woosley_02} in 2D and 3D, using the 
\textsc{Prometheus-Vertex} code with detailed multi-group neutrino
transport including the full, sophisticated set of neutrino
reactions applied previously in 1D and 2D supernova modeling by
the Garching group. Moreover, we performed a set of 2D and 3D 
post-bounce simulations of the 25\,$M_\odot$ progenitor of
\citet{woosley_02} with the \textsc{Prometheus} hydrodynamics
scheme and a computationally efficient, gray neutrino transport
approximation, employing a parametric approach in which an inner
grid boundary replaced the excised high-density core of the 
proto-neutron star and allowed us to more systematically 
explore the influence of different core neutrino luminosities 
and of faster or slower contraction of the forming remnant.
We emphasize that this approach is different from that for
our 27\,$M_\odot$ runs, where the whole neutron
star to the center was included in the computational domain.
In addition, we also tested the effects of varied amplitudes
of random perturbations that had to be imposed for seeding the
growth of nonradial hydrodynamic instabilities
in the neutrino-heated accretion layer behind the stalled 
supernova shock. While the 3D run for the 27\,$M_\odot$ star
was conducted with a polar coordinate grid, an axis-free 
Yin-Yang grid was used in the studies of the 25\,$M_\odot$
progenitor.

Our simulations for the first time provide unambiguous evidence of the
occurrence of large-amplitude SASI shock sloshing and spiral motions
and of their interplay with neutrino-driven convection in 3D supernova
core environments modeled with a ``realistic'' treatment of neutrino
transport and the corresponding heating and cooling effects. Previous
hydrodynamic studies had been able to identify SASI activity, and in
particular SASI spiral modes, only in 3D setups with adiabatic
postshock accretion flows \citep{blondin_07,fernandez_10} and in some
simulations with a simple neutrino light-bulb treatment
\citep{iwakami_08,iwakami_09}.  The shallow water analogue of the SASI
was also observed experimentally, however again without the presence
of buoyancy motions \citep{foglizzo_12}. On the basis of more recent
3D simulations with a simple neutrino light-bulb approximation
\citep{burrows_12,murphy_12,dolence_13} and 3D GR models with a
neutrino leakage scheme \citep{ott_12} it was even concluded that the
SASI is at most of minor relevance for the dynamics of the postshock
layer in collapsing stellar cores (see also \citealp{burrows_13}).

Our findings indicate that the SASI is potentially much more 
important in 3D than suggested by these previous investigations.
Besides the core-density profile of the progenitor star, which
determines the temporal evolution of the mass infall rate, the
more elaborate neutrino transport treatment (in particular in our
27\,$M_\odot$ simulation) may partly be responsible for the
different accretion-flow dynamics seen in our models. The
details of the neutrino description are relevant, because they
can cause differences in the contraction behavior of the 
proto-neutron star and in the neutrino-heating in the gain layer.
Both affect the evolution of the stagnation radius of the 
accretion shock and thus have a strong impact on the growth
conditions for convection and the SASI. In the artificial
setup used by \citet{burrows_12,murphy_12,dolence_13}, the
neutron star is not allowed to deleptonize and radiate energy,
therefore its radius stays unrealistically large (50--60\,km) 
during the whole simulated postbounce evolution (up to $\sim$1\,s).
Also the leakage scheme of \citet{ott_12},
which can track the contraction of the proto-neutron star
and the time evolution of the neutrino emission to a certain
extent, deviates significantly from full transport models. 
For example, in the work of \citet{ott_12} the mean energy of 
the radiated electron neutrinos around 100\,ms after core bounce
is more than 60\% higher
than that found with the \textsc{Vertex} transport code by
\citet{mueller_12b}, although both groups investigated the same
27\,$M_\odot$ progenitor with relativistic methods. It is clear
that the neutrino heating found by \citet{ott_12} was stronger
and thus more favorable for larger shock radii, which is turn
was supportive of neutrino-driven convection instead of the 
SASI.

In contrast to the findings by
\citet{burrows_12,murphy_12,dolence_13} and by \citet{ott_12},
we observe strong SASI activity in our 3D simulations both for 
the 27\,$M_\odot$ progenitor and the 25\,$M_\odot$ star. The SASI 
can become the clearly strongest nonradial instability during at
least some parts of the evolution and can be clearly identified
by its oscillatory growth properties and even in the nonlinear regime 
by its characteristic low-order spherical harmonics modes of coherent
shock motion and shock deformation. In detail, we can draw the 
following conclusions from our models:
\begin{itemize}
\item 
SASI activity can develop in 3D despite and in addition to 
the earlier presence of
neutrino-driven buoyancy. Different from the higher-order 
multipole pattern that is typical of convective plumes and
downdrafts, the SASI reveals itself by coherent,
large-amplitude shock sloshing and spiral motions.
\item
Because of the absence of a flow-constraining symmetry axis,
which also directs the structure of neutrino-driven buoyancy
in 2D models, SASI shock motions and deformation modes can be
recognized more easily and more clearly in the 3D case.
Interestingly, both the 27\,$M_\odot$ and 25\,$M_\odot$ models
exhibit evolution phases in which the SASI in 3D can become
stronger than in the corresponding 2D runs. The dominance of 
the SASI and greater strength in 3D 
can be concluded not only from the large 
dipole and quadrupole amplitudes of the shock deformation, but
also from the higher nonradial kinetic energy of the postshock
flow (Figure~\ref{fig:overview}) and from a prominent
peak at low-order multipoles in the power spectrum of the 
nonradial kinetic energy (Figure~\ref{fig:spectrum}).
While some authors
hypothesized \citep{iwakami_08} that the SASI 
amplitudes in 3D remain smaller than those in 2D because the 
kinetic energy of the nonradial flow is shared with an additional
degree of freedom, our results suggest that there is no such 
limitation of the kinetic energy that can be stored in lateral
and azimuthal mass motions. The fraction of the huge reservoir
of accretion energy that is converted to nonradial flows in the 
postshock layer can be larger in 3D than in 2D.
\item
Depending on the conditions in the postshock flow, which in our
25\,$M_\odot$ runs could be controlled by the choices of the 
contraction behavior and neutrino luminosities assumed at the inner 
grid boundary, phases of essentially {\em pure} SASI activity
could be obtained, with neutrino-driven convection  only
developing as the secondary instability (cf.\ Figure~\ref{fig:2d3dsasi},
middle panels). This result of the 2D and 3D models of the present
work is very similar to the flow behavior seen in the 2D 
simulation of the 27\,$M_\odot$ progenitor by
\citet{mueller_12b}.
\item
The growth of the SASI is favored by fast advection flows, because
its growth rate in an advective-acoustic cycle scales roughly 
inversely with the advection timescale of the accretion flow from
the shock to the neutron star surface (e.g., \citealp{scheck_08}). 
This is opposite to neutrino-driven buoyancy, whose growth
in the accretion flow requires that 
the critical ratio $\chi$ of advection
timescale to buoyancy timescale exceeds a value of about 3 
\citep{foglizzo_06}. Fully consistent with this dimension-independent 
theoretical understanding, which was developed by linear analysis
and hydrodynamical modeling in 2D, we find SASI growth also in the 
3D case to be strongest in phases of small shock stagnation radii,
which are connected to rapidly shrinking and more compact 
proto-neutron stars as well as relatively weak neutrino heating.
\item
Also the amplitude of the initial seed perturbations (of the density
or velocity field) is found to potentially have an influence.
A greater amplitude can trigger a faster
development of neutrino-driven convection (for perturbations of 
nonlinear size this can happen
even when $\chi < 3$, see \citealt{scheck_08}).
If the buoyancy instability is supported by sufficiently strong
neutrino heating to instigate an expansion of the shock, SASI
growth can become disfavored and nonradial flows in the
postshock layer are dominated by buoyant plumes and downflows.
\item
Preferentially in phases of strongest shock recession we observe
bipolar ($\ell = 1$, $m = 0$) SASI sloshing motions to change 
over to a spiral ($\ell = 1$, $m = 1$) mode in the 3D
simulations of both the 27\,$M_\odot$ and 25\,$M_\odot$ models
(Figures~\ref{fig:3d_amplitude} and \ref{fig:spiral}).
The transition to the time-dependent rotating shock-deformation 
pattern therefore seems to be favored by particularly
small shock radii. In general, the character of the 3D 
accretion flow during shock oscillation and spiraling phases with
wandering directions reveals close similarity to the behavior of
the hydraulic jump observed in the SWASI experiment 
\citep{foglizzo_12}.
\end{itemize}

Our comparison of the SASI activity in the 2D and 3D simulations of 
the 27\,$M_\odot$ model, which was evolved fully self-consistently 
(i.e., without the use of free parameters and without an inner grid 
boundary) is particularly interesting. It revealed that at an early 
stage, when the mass accretion rate is still high and the shock 
correspondingly retreats in response to the proto-neutron star
contraction, the SASI can grow
despite some prior neutrino-driven convective activity. Before the
accretion of the Si/SiO interface, the SASI can reach even higher 
amplitudes in 3D than in 2D. The large-amplitude shock-sloshing mode 
eventually makes the transition to a spiral mode in 3D, which, however,
is quenched around $260 \ \mathrm{ms}$ after bounce once the accretion
rate has dropped significantly and the shock has expanded to radii
of nearly 200\,km on average. Despite the faster shock expansion and
higher kinetic energy of the postshock flow, the 3D model nevertheless
falls short of an explosive runaway unlike the 2D simulation.

A discussion of this important difference in the explosion behavior of 
2D and 3D models, which is fully consistent with the results obtained 
in the simulations with simple neutrino heating and 
cooling treatments of \citet{hanke_12} and \citet{couch_12},
is beyond the scope of this paper and 
will be addressed in future work. We do not want to speculate here 
about the relevance of the SASI in competition with neutrino-driven 
convection for getting explosions in 3D. It is possible that the SASI
could provide crucial support for the onset of the explosion, 
especially since a very strong spiral mode returns towards the end
of our 27\,$M_\odot$ simulation. Such a SASI supported explosion was
obtained in our parametric 25\,$M_\odot$ simulation of model S25-2
(Figure~\ref{fig:2d3dcomp}, middle column). 
But it is also possible that its presence is only temporary and
that neutrino-driven convection dominates when the explosion sets
in and the preceding shock expansion leads to disfavorable 
conditions for the advective-acoustic cycle feeding the SASI.
Even in the latter case, the SASI could still
play an essential role in the explosion mechanism by
pushing out the shock far enough for convection to take over
in cases where convection cannot develop on its own in an initially
stabilized post-shock flow.
The conditions for the growth and long-time persistence of the SASI,
in particular of the spiral mode, are still poorly understood and
have to be explored in more detail by future 3D simulations. Our
results suggest that the exact role of the SASI will depend strongly
and in a subtle way not only on a variety of progenitor dependent 
conditions like the steepness of the density profile, the location 
and sharpness of the composition shell interfaces, the existence of 
initial nonradial asymmetries in the collapsing stellar core, 
angular momentum of the progenitor core, which could trigger a 
faster growth of the spiral mode even for small rotation rates
\citep{yamasaki_08}, and magnetic fields, if
these are amplified to dynamically relevant strength.
The SASI growth conditions will also depend on the contraction
behavior of the nascent neutron star, which is influenced by its 
high-density equation of state and by general relativistic gravity,
and on the strength of the neutrino heating, whose determination
requires a sophisticated treatment of the energy-dependent neutrino 
transport. 

The initial perturbations constitute a potential cause for
differences that is very difficult to address. Certainly neither
the artificially imposed random perturbations in our models, nor the
small intrinsic numerical seeds of \citet{mueller_12b}, nor the larger
intrinsic perturbations in the simulations of \citet{ott_12} reflect
the magnitude and spectral shape of the physical perturbations present
in realistic progenitors. In the end, a well-founded understanding
of the interplay between convection and the SASI will probably require
multi-dimensional progenitor models.

If we suppose that the growth conditions for the SASI remain favorable
for realistic seed perturbations, our self-consistent 3D model of the 
27\,$M_\odot$ progenitor still throws up a
number of further questions: Has the SASI already reached its
saturation level in our model, or could it grow further if massive
accretion were to last longer? Could it play a decisive dynamical role
for reaching an explosive runaway in the latter case, even if it is
only by pushing the shock out far enough for convection to take over
later? Are the saturation properties of the nonlinear spiral mode
different from the nonlinear sloshing mode because of the different
geometry of the downflows and the presence of considerable angular
momentum in the gain region? If the strong and clear spiral mode
survives until the onset of an explosion, what would be its final
impact on the proto-neutron star kick and spin? 
What are the neutrino-emission variations
connected to large-amplitude shock sloshing and spiral motions
in the 3D case? Are these signal variations observable with a 
similar strength as predicted for the 2D case
(see \citealp{marek_09,lund_10})?
With sophisticated 3D supernova modeling only now beginning to 
become feasible, our understanding of 3D
effects in supernova cores is still in its infancy, and
further research into the properties of the SASI is urgently
needed.

\acknowledgements 
We are grateful to Elena Erastova and Markus Rampp 
(Rechenzentrum Garching) for visualizing the 3D simulation in
Figure~\ref{fig:3dsasi}. This research was supported by the Deutsche
Forschungsgemeinschaft through the Transregional Collaborative
Research Center SFB/TR 7 ``Gravitational Wave Astronomy'' and the
Cluster of Excellence EXC 153 ``Origin and Structure of the Universe''
(http://www.universe-cluster.de). 
The results described in this paper could only be
achieved with the assistance of high performance computing resources 
(Tier-0) provided by PRACE on CURIE TN (GENCI@CEA, France) and 
SuperMUC (GCS@LRZ, Germany). We also thank the Rechenzentrum
Garching for computing time on the IBM iDataPlex system \emph{hydra}.

\bibliography{paper}

\end{document}